\documentstyle[epsfig,12pt,graphicx]{article}
\newcommand{\prepr}[1] {\begin{flushright}  {\bf #1} \end{flushright}
\vskip 1.cm}
\newcommand{\titul}[1] {\begin{center}{\Large {\bf #1 } } \end{center}
\vskip 0.8cm}

\newcommand{\autor}[1] {\begin{center}  {\bf \lineskip .3cm #1  }
                        \end{center} }

\newcommand{\lugar}[1] {\begin{center}  {\normalsize \bf \it #1   }
\end{center}}
\begin{document}
\hbadness=10000
\pagenumbering{arabic}
\begin{titlepage}
\prepr{ \hspace{20mm} TU-735}
\titul{\bf Polarizations in $B\to VV$ decays}
\autor{Hsiang-nan Li$^{1}$\footnote{E-mail:
hnli@phys.sinica.edu.tw} and Satoshi Mishima$^2$\footnote{E-mail:
mishima@tuhep.phys.tohoku.ac.jp} }

\lugar{$^{1}$Institute of Physics, Academia Sinica, Taipei, Taiwan
115, Republic of China}
\lugar{$^{1}$Department of Physics, National Cheng-Kung University,\\
Tainan, Taiwan 701, Republic of China}

\lugar{$^{2}$Department of Physics, Tohoku University, Sendai
980-8578, Japan}

\vskip 2.0cm {\bf  PACS index : 13.25.Hw, 11.10.Hi, 12.38.Bx}

\thispagestyle{empty}
\vspace{10mm}
\begin{abstract}

We demonstrate that the polarization fractions of most
tree-dominated $B\to VV$ decays can be simply understood by means
of kinematics in the heavy-quark or large-energy limit. For
example, the longitudinal polarization fractions $R_L$ of the
$B^0\to (D_s^{*+}, D^{*+}, \rho^+)D^{*-}$ and $B^+\to (D_s^{*+},
D^{*+}, \rho^+)\rho^0$ modes increase as the masses of the mesons
$D_s^{*+}, D^{*+}, \rho^+$ emitted from the weak vertex decrease.
The subleading finite-mass or finite-energy corrections modify
these simple estimates only slightly. Our predictions for the
$B\to D_{(s)}^* D^*$ polarization fractions derived in the
perturbative QCD framework, especially $R_L\sim 1$ for $B^0\to
{\bar D}^{*0} D^{*0}$ governed by nonfactorizable $W$-exchange
amplitudes, can be confronted with future data. For
penguin-dominated modes, such as $B\to\rho(\omega) K^*$, the
polarization fractions can be understood by the annihilation
effect from the $(S-P)(S+P)$ operators, plus the interference with
a small tree amplitude. At last, we comment on the various
mechanism proposed in the literature to explain the abnormal $B\to
\phi K^*$ polarization data, none of which are satisfactory.

\end{abstract}
\thispagestyle{empty}
\end{titlepage}


\section{INTRODUCTION}

The measured polarization fractions in the $B\to\phi K^*$ decays
have exhibited an anomaly. Let $R_{L,\parallel,\perp}$ denote the
longitudinal, parallel, and perpendicular polarization fractions
of a $B\to VV$ mode, respectively. It is well known from a
helicity argument that these fractions for light vector mesons $V$
follow the naive counting rules,
\begin{eqnarray}
R_L\sim 1-O(m_V^2/m_B^2)\;,\;\;\;\;R_\parallel \sim R_\perp\sim
O(m_V^2/m_B^2)\;, \label{nai}
\end{eqnarray}
if the emission topology of diagrams dominates, where $m_B$
($m_V$) is the $B$ ($V$) meson mass. That is, Eq.~(\ref{nai})
holds for tree-dominated modes, such as $B \to \rho \rho$, whose
longitudinal polarization fraction has been observed to be
$R_L\sim 1$ \cite{BelleRhopRho0,BaBarRhopRho0}. For
penguin-dominated modes, Eq.~(\ref{nai}) could be modified by
annihilation contributions to some extent \cite{CKL2,Li04}.
However, the $B \to \phi K^{*}$ polarization fractions shown in
Table~\ref{tab:tab1} were found to dramatically differ from the
naive counting rules, and have been considered as a puzzle.

\begin{table}[ht]
\begin{center}
\begin{tabular}{c c c c}\hline \hline
Mode&Pol. Fraction&Belle&Babar \\
\hline
$B^+\to\phi K^{*+}$&$R_L$&$0.49\pm 0.13\pm 0.05$ \cite{Zhang04}&$0.46\pm0.12\pm0.03$ \cite{BaBarRhopRho0}\\
&$R_\perp$&$0.12^{+0.11}_{-0.08}\pm 0.03$ \cite{Zhang04}&\\
$B^0\to\phi K^{*0}$&$R_L$&$0.52\pm 0.07\pm 0.05$ \cite{Zhang04} 
&$0.52\pm0.05\pm0.02$ \cite{Bar017}\\
&$R_\perp$&$0.30\pm 0.07\pm 0.03$ \cite{Zhang04}
&$0.22\pm0.05\pm0.02$ \cite{Bar017}\\
\hline \hline
Mode&Pol. Fraction&Belle&Babar\\
\hline
$B^+\to\rho^0 K^{*+}$&$R_L$&&$0.96^{+0.04}_{-0.15}\pm0.04$ \cite{BaBarRhopRho0}\\
$B^+\to\rho^+ K^{*0}$&$R_L$&$0.50\pm 0.19^{+0.05}_{-0.07}$ \cite{Bel102}&$0.79\pm 0.08\pm0.04\pm 0.02$ \cite{Bar093}\\
\hline \hline
\end{tabular}
\end{center}
\caption{Polarization fractions in the penguin-dominated $B\to VV$
decays. }\label{tab:tab1}
\end{table}

In this work we shall investigate all the $B\to VV$ polarization
data carefully, and understand more the above puzzle. We show that
the $B\to VV$ modes can be classified into four categories: the
first category can be easily understood by means of kinematics in
the heavy-quark limit. Take the $B^0\to (D_s^{*+}, D^{*+},
\rho^+)D^{*-}$ modes as examples. Under the naive factorization
assumption (FA) \cite{BSW}, QCD dynamics in different helicity
amplitudes is absorbed into the universal Isgur-Wise (IW) function
\cite{IW}. The polarization fractions are then completely
determined by the kinematic factors, leading to the observation
that $R_L$ increases from 0.5 to 0.9, when the vector mesons from
the weak vertex change through $D_s^{*}$, $D^{*}$ and $\rho$. We
shall examine subleading effects by deriving the perturbative QCD
(PQCD) \cite{KLS,LUY} factorization formulas for the $B\to
D_{(s)}^*D^*$ decays up to next-to-leading power in
$m_{D_{(s)}^*}/m_B$, $m_{D_{(s)}^*}$ being the $D_{(s)}^*$ meson
mass. To this level of accuracy, the various form factors deviate
from the IW function, and the nonfactorizable contributions
appear. It will be demonstrated that the simple kinematic
estimates are robust under these subleading corrections. As a
byproduct, we observe that $R_L\sim 1$ for $B^0\to {\bar D}^{*0}
D^{*0}$, governed by nonfactorizable $W$-exchange amplitudes,
differs from $R_L\sim 0.5$ for other $B\to D^* D^*$ modes. This
PQCD prediction can be confronted with data in the future.

The second category is understandable via kinematics in the
large-energy limit, which consists of tree-dominated $B$ meson
decays into light vector mesons, such as $B\to (\rho,
\omega)\rho$. Their helicity amplitudes can be expressed in terms
of various heavy-to-light transition form factors under FA, which
are related to each other by the large-energy symmetry relations.
It turns out that only two universal form factors associated with
the longitudinally and transversely polarized final states are
relevant. If these two form factors do not differ much, the
polarization fractions will be completely determined by the
kinematic factors, leading to $R_L\sim 1$ consistent with
Eq.~(\ref{nai}). The same argument applies to the $B^+\to
(D_s^{*+}, D^{*+})\rho^0$ decays with the $D_s^{*+}, D^{*+}$
mesons being emitted from the weak vertex, whose $R_L\sim 0.7$ are
predicted, and can be compared with the future data. We then
examine subleading corrections to the large-energy symmetry
relations by including the two-parton twist-4 contribution. Adding
this piece to the longitudinal polarization amplitude makes
complete the next-to-leading-power analysis at two-parton level,
since the transverse polarization amplitudes are of
next-to-leading power by themselves. It will be demonstrated that
this subleading effect is also negligible.

The third category contains penguin-dominated modes, such as
$B\to\rho (\omega) K^*$, which can exhibit a sizable deviation
from the naive counting rules in Eq.~(\ref{nai}). As shown in
\cite{CKL2}, the annihilation amplitudes associated with the
$(S-P)(S+P)$ operators follow the counting rules,
\begin{eqnarray}
R_L\sim R_\parallel \sim R_\perp\;. \label{mod}
\end{eqnarray}
The PQCD analysis has indicated that the penguin annihilation
contribution, together with nonfactorizable corrections, bring the
longitudinal polarization fraction in a pure-penguin mode from
$R_L\sim 0.9$ down to $0.75$ \cite{CKL2}. Therefore, the
$B^+\to\rho^+K^{*0}$ polarization data in Table~\ref{tab:tab1} can
be well accommodated within the Standard Model, showing no
anomaly. The longitudinal polarization fraction of another mode
$B^+\to\rho^0K^{*+}$ remains as $R_L\sim 0.9$ because of the
interference between the penguin amplitude and an additional tree
amplitude. From the viewpoint of PQCD, the $B\to\omega K^*$ decays
do not differ much from $B\to\rho^0 K^*$, and are expected to show
similar $R_L$. This prediction can be tested by the future data.

The fourth category, involving the $B\to\phi K^*$ decays, is the
abnormal one. These decays occur mainly through the penguin
operators, but their $R_L$ are as small as 0.5, much lower than
0.75 expected from PQCD. The mechanism proposed in the literature
to explain the $B\to\phi K^*$ polarization data includes new
physics \cite{G03}, the annihilation contribution \cite{AK} in the
framework based on the QCD-improved factorization (QCDF)
\cite{BBNS}, the charming penguin \cite{BPRS}, the rescattering
effect \cite{CDP,LLNS,CCS}, and the $b\to sg$ transition (the
magnetic penguin) \cite{HN}. We shall comment on these proposals:
the annihilation contribution has to be parameterized in QCDF, and
varying free parameters to fit the data can not be conclusive. The
charming penguin strategy, demanding many free parameters, does
not help understand dynamics. The rescattering effect is based on
a model-dependent analysis \cite{W,Ligeti04}, and constrained by
the $B\to\rho K^{*}$ data. The magnetic penguin is suppressed by
the $G$-parity, and not sufficient to reduce $R_L$ down to 0.5.
Therefore, none of these proposals is satisfactory \cite{Li04}.
However, we are not concluding that the $B\to\phi K^*$
polarization data signal new physics, since the complicated QCD
dynamics in $B\to VV$ modes has not yet been fully explored.

In Sec.~II we study the kinematic effects on the polarizations of
tree-dominated decays using FA in the heavy-quark or large-energy
limit. The next-to-leading-power corrections to the $B\to
D_{(s)}^*D^*$ decays, and the two-parton twist-4 corrections to
the $B\to\rho\rho$ decays are calculated in Sec.~III. We comment
on the proposals for explaining the abnormal $B\to\phi K^*$ data
in Sec.~IV. Section V is the conclusion.

\section{NAIVE FACTORIZATION}

In this section we demonstrate that the polarization fractions of
tree-dominated $B\to VV$ decays can be simply understood by means
of kinematics in the heavy-quark or large-energy limit.

\subsection{Heavy-quark Limit}

We first investigate the polarizations in the decays $B^0\to
(D_s^{*+}, D^{*+}, \rho^+)D^{*-}$ in the heavy quark limit. These
modes are color-allowed with the $D_s^{*+}, D^{*+}, \rho^+$ mesons
emitted from the weak vertex, respectively, and FA is supposed to
work well. Take the $B^0\to D_s^{*+} D^{*-}$ decay as an example.
The $B$ meson momentum $P_1$, the $D^*$ meson momentum $P_2$, and
the $D_s^*$ meson momentum $P_3$ are chosen, in the light-cone
coordinates, as
\begin{eqnarray}
P_1&=&\frac{m_B}{\sqrt{2}}(1,1,0_T)\;,\nonumber\\
P_2&=&\frac{m_B}{\sqrt{2}}(r_2\eta^+,r_2\eta^-,{\bf 0}_T)\;,\nonumber\\
P_3&=&\frac{m_B}{\sqrt{2}}(1-r_2\eta^+,1-r_2\eta^-,{\bf 0}_T)\;,
\label{pal}
\end{eqnarray}
where the factors $\eta^\pm$ are defined by
$\eta^\pm=\eta\pm\sqrt{\eta^2-1}$ with $\eta=v_1\cdot v_2$ being
the velocity transfer, $v_1\equiv P_1/m_B$ and $v_2\equiv
P_2/m_{D^*}$ the $B$ meson and $D^*$ meson velocities,
respectively, and $r_2=m_{D^*}/m_B$ the mass ratio. To extract the
helicity amplitudes, the following parametrization for the
longitudinal polarization vectors is useful:
\begin{eqnarray}
\epsilon_2(L)&=&v_2-\frac{m_{D^*}}{P_2\cdot
n_-}n_-=\frac{1}{\sqrt{2}}(\eta^+,-\eta^-,{\bf 0}_T)\;,\nonumber\\
\epsilon_3(L)&=&v_3-\frac{m_{D_s^*}}{P_3\cdot
n_+}n_+=\frac{1}{\sqrt{2}r_3}\left(-\frac{r_3^2}{1-r_2\eta^-},1-r_2\eta^-,{\bf
0}_T\right)\;,\label{lpo}
\end{eqnarray}
with the $D_s^*$ meson velocity $v_3\equiv P_3/m_{D_s^*}$, the
mass ratio $r_3=m_{D_s^*}/m_B$, and the null vectors
$n_+=(1,0,{\bf 0}_T)$ and $n_-=(0,1,{\bf 0}_T)$, which satisfy the
normalization $\epsilon_2^2(L)=\epsilon_3^2(L)=-1$ and the
orthogonality $\epsilon_2(L)\cdot P_2=\epsilon_3(L)\cdot P_3=0$.
For the transverse polarization vectors, we simply choose
\begin{eqnarray}
\epsilon_2(T) =(0,0,{\bf 1}_T)\;,\;\;\;\; \epsilon_3(T) =(0,0,{\bf
1}_T)\;.\label{tpo}
\end{eqnarray}

The relevant effective weak Hamiltonian is given by
\begin{eqnarray}
{\cal H}_{\rm eff} = {G_F\over\sqrt{2}}\, V_{cb}V_{cs}^*
\Big[C_1(\mu)O_1(\mu)+C_2(\mu)O_2(\mu)\Big]\;,
\end{eqnarray}
where $V$'s are the Cabibbo-Kobayashi-Maskawa (CKM) matrix
elements, $C_1$ and $C_2$ the Wilson coefficients, and
\begin{eqnarray}
O_1= (\bar sb)_{V-A}(\bar cc)_{V-A}\;,\qquad\qquad O_2= (\bar
cb)_{V-A}(\bar sc)_{V-A}\;,
\end{eqnarray}
the four-fermion operators with the definition $(\bar
q_1q_2)_{V-A}\equiv \bar q_1\gamma_\mu(1- \gamma_5)q_2$. The
contributions from $O_1$ and $O_2$ can be combined, and the
resultant coefficient appears as $a_1= C_2 + C_1/ N_c$, $N_c$
being the number of colors.

The $B^0\to D_s^{*+} D^{*-}$ decay amplitude in FA is expressed as
\begin{eqnarray}
{\cal M}^{(\sigma)}&=&\langle D_s^{*+}(P_3,\epsilon_3^*)
D^{*-}(P_2,\epsilon_2^*) | {\cal H}_{\rm eff} | B^0(P_1)
\rangle\;,\nonumber\\
&=& \displaystyle{G_F \over \sqrt{2}}V_{cb}^*V_{cs} a_1 \langle
D_s^{*+}(P_3,\epsilon_3^*)|{\bar c}\gamma_\mu(1-\gamma_5) s|0
\rangle\nonumber\\
& &\times\langle D^{*-}(P_2,\epsilon_2^*)|{\bar
b}\gamma^\mu(1-\gamma_5) c|B^0(P_1) \rangle\;, \label{nfa}
\end{eqnarray}
where the superscript $\sigma$ denotes a possible final helicity
state. The first matrix element defines the $D_s^*$ meson decay
constant,
\begin{eqnarray}
\langle D_s^{*+}(P_3,\epsilon_3^*)|{\bar c}\gamma_\mu(1-\gamma_5)
s|0 \rangle=m_{D_s^*}f_{D_s^*}\epsilon_{3\mu}^*\;.\label{dcn}
\end{eqnarray}
The matrix elements for the $B\to D^*$ transitions are
parameterized as
\begin{eqnarray}
\langle D^{*-}(P_2,\epsilon_2^*)|{\bar b}\gamma^\mu
c|B^0(P_1)\rangle&=&
i\sqrt{m_Bm_{D^*}}\xi_V(\eta)\epsilon^{\mu\nu\alpha\beta}
\epsilon^*_{2\nu} v_{2\alpha}v_{1\beta}\;,\nonumber\\
\langle D^{*-}(P_2,\epsilon_2^*)|{\bar b}\gamma^\mu\gamma_5
c|B^0(P_1) \rangle&=&\sqrt{m_Bm_{D^*}}
\left[{\xi_{A1}}(\eta)(\eta+1)\epsilon^{*\mu}_2-{\xi_{A2}}(\eta)
\epsilon_2^*\cdot v_1v_1^\mu\right.\nonumber\\
& &\left.-{\xi_{A3}}(\eta)\epsilon_2^*\cdot v_1 v_2^\mu \right]
\;, \label{bds}
\end{eqnarray}
where the form factors $\xi_{A_1}$, $\xi_{A_2}$, $\xi_{A_3}$, and
$\xi_V$ satisfy the relations in the heavy-quark limit,
\begin{equation}
\xi_V=\xi_{A_1}=\xi_{A_3}=\xi,\;\;\;\;  \xi_{A_2}=0\;, \label{iwr}
\end{equation}
with $\xi$ being the IW function \cite{IW}.

The $B^0\to D_s^{*+} D^{*-}$ decay rate is given by
\begin{equation}
\Gamma =\frac{P_c}{8\pi m^{2}_{B} }\sum_\sigma
{\cal M}^{(\sigma)\dagger }{\cal M^{(\sigma)}}\;,
\label{dr1}
\end{equation}
where $P_c\equiv |P_{2z}|=|P_{3z}|=m_Br_2\sqrt{\eta^2-1}$
is the momentum of either of the vector mesons. The amplitude
$\cal M^{(\sigma)}$ is decomposed into
\begin{eqnarray}
{\cal M}^{(\sigma)}
=\left(m_{B}^{2}{\cal M}_{L}, m_{B}^{2}{\cal M}_{N}
\epsilon^{*}_{2}(T)\cdot\epsilon^{*}_{3}(T), i{\cal
M}_{T}\epsilon^{\alpha \beta\gamma \rho}
\epsilon^{*}_{2\alpha}\epsilon^{*}_{3\beta} P_{2\gamma }P_{3\rho
}\right)\;,
\end{eqnarray}
where the first term corresponds to the configuration with both
the vector mesons being longitudinally polarized, and the second
(third) term to the two configurations with both the vector mesons
being transversely polarized in the parallel (perpendicular)
directions. The helicity amplitudes are then defined as,
\begin{eqnarray}
A_{L}&=&-G m^{2}_{B}{\cal M}_{L}, \nonumber\\
A_{\parallel}&=&G \sqrt{2}m^{2}_{B}{\cal M}_{N}, \nonumber \\
A_{\perp}&=&G m_{D_s^*} m_{D^*} \sqrt{2[(v_2\cdot v_3)^{2}-1]}
{\cal M }_{T}\;, \label{ase}
\end{eqnarray}
with the normalization factor $G=\sqrt{P_c/(8\pi m^2_{B}\Gamma)}$,
which satisfy the relation,
\begin{eqnarray}
|A_{L}|^2+|A_{\parallel}|^2+|A_{\perp}|^2=1\;.
\end{eqnarray}
It is easy to read the helicity amplitudes off
Eqs.~(\ref{nfa})-(\ref{bds}),
\begin{eqnarray}
A_{L}&\propto&
\epsilon_2^*(L)\cdot \epsilon_3^*(L)(\eta+1)\xi_{A_1}
-\epsilon_2^*(L)\cdot v_1\left[\epsilon_3^*(L)\cdot v_1\xi_{A_2}
+\epsilon_3^*(L)\cdot v_2\xi_{A_3}\right]
\;, \nonumber\\
A_{\parallel}&\propto&- \sqrt{2}
(\eta+1)\xi_{A_1}\;, \nonumber \\
A_{\perp}&\propto&-
r_3\sqrt{2[(v_2\cdot v_3)^{2}-1]}\xi_V\;. \label{ase1}
\end{eqnarray}

In the heavy-quark limit, i.e., applying Eq.~(\ref{iwr}), all the
helicity amplitudes depend on a single IW function $\xi$, which
absorbs QCD dynamics.
Simply inserting $m_B=5.28$ GeV, $m_{D_s^*}=2.11$ GeV, and
$m_{D^*}= 2.01$ GeV into the kinematic factors, we obtain
\begin{eqnarray}
R_L\sim 0.52\;,\;\;\;\; R_{\parallel}\sim 0.43\;,\;\;\;\;
R_{\perp}\sim 0.05\;. \label{arr}
\end{eqnarray}
Equation (\ref{ase1}) is applicable to the $B^0\to D^{*+}D^{*-}$
and $B^0\to \rho^{+}D^{*-}$ decays by substituting $m_{D^*}$ and
$m_\rho=0.77$ GeV for $m_{D_s^*}$, leading to
\begin{eqnarray}
& &R_L\sim 0.54\;,\;\;\;\; R_{\parallel}\sim 0.41\;,\;\;\;\;
R_{\perp}\sim 0.05\;, \label{ard}\\
& &R_L\sim 0.88\;,\;\;\;\; R_{\parallel}\sim 0.10\;,\;\;\;\;
R_{\perp}\sim 0.02\;, \label{ar}
\end{eqnarray}
respectively. All the above results are consistent with the
observed values listed in Table~\ref{tab:tab2}, and with the
predictions in \cite{Ros90}. The estimated polarization fractions
for the decay $B^-\to K^{*-}D^{*0}$ are close to Eq.~(\ref{ar})
and consistent with the data. For the $B^+\to \rho^{+}\bar D^{*0}$
decay, the internal-$W$ emission amplitudes involving the
$B\to\rho$ form factors are suppressed by the vanishing
coefficient $a_2=C_1+C_2/N_c$. Hence, this mode is similar to
$B^0\to \rho^{+} D^{*-}$, and the result in Eq.~(\ref{ar}) applies
as shown in Table~\ref{tab:tab2}. It is easy to find that the
longitudinal polarization fraction increases as the mass of the
vector meson from the weak vertex decreases.

\begin{table}[ht]
\begin{center}

\begin{tabular}{c c c c}\hline \hline
Mode
& Pol. Fraction& Data\\
\hline
$B^0\to D_s^{*+} D^{*-}$&$R_L$
&$0.52\pm0.05$ \cite{PDG}\\
$B^0\to D^{*+} D^{*-}$ &$R_L$
&$0.57\pm0.08\pm0.02$ \cite{Belich}\\
&$R_\perp$ &$0.19\pm0.08\pm0.01$ \cite{Belich}\\
&$R_\perp$ & $0.063\pm 0.055\pm 0.009$ \cite{PDG,BAR0306}\\
$B^0\to \rho^{+} D^{*-}$&$R_L$
&$0.885\pm0.016\pm0.012$ \cite{PDG}\\
$B^+\to \rho^{+} {\bar D}^{*0}$&$R_L$
&$0.892\pm0.018\pm0.016$ \cite{PDG}\\
$B^-\to K^{*-} D^{*0}$ &$R_L$
&$0.86\pm0.06\pm0.03$ \cite{BAR0308}\\
\hline \hline
Mode& Pol. Fraction &Belle &Babar\\
\hline
$B^+\to\rho^+ \rho^0$&$R_L$&$0.95\pm0.11\pm0.02$ \cite{BelleRhopRho0}&$0.97^{+0.03}_{-0.07}\pm0.04$ \cite{BaBarRhopRho0}\\
$B^0\to\rho^+ \rho^-$&$R_L$&&$0.99 \pm 0.03^{+0.04}_{-0.03}$ \cite{Aubert:2003xc}\\
$B^+\to \rho^+\omega$&$R_L$&&$0.88^{+0.12}_{-0.15}\pm 0.03$ \cite{gritsan}\\
\hline \hline
\end{tabular}
\end{center}
\caption{Polarization fractions in the tree-dominated $B\to VV$
decays.}\label{tab:tab2}
\end{table}

\subsection{Large-energy Limit}

We then show that the polarization fractions in the $B\to
(\rho,\omega)\rho$ decays can be understood by means of kinematics
in the large-energy limit. The parametrizations of the momenta in
Eq.~(\ref{pal}) and of the polarization vectors in
Eqs.~(\ref{lpo}) and (\ref{tpo}) hold here, with the mass ratio
$r_2$ for the vector meson from the $B$ meson transition and $r_3$
for the vector meson emitted from the weak vertex. The transition
form factors associated with a $B\to V$ transition are defined via
the matrix elements,
\begin{eqnarray}
\langle V(P_2,\epsilon_2^\ast)| \bar b \gamma^\mu q |B(P_1)
\rangle& =&
 \frac{2iV(q^2)}{m_B+m_V} \epsilon^{\mu\nu\rho\sigma}
 \epsilon_{2\nu}^{\ast}  P_{2\rho} P_{1\sigma},
\label{V}\\
\langle V(P_2,\epsilon_2^\ast)|\bar b \gamma^\mu\gamma_5 q |
B(P_1) \rangle &=&
  2m_VA_0(q^2)\frac{\epsilon_2^\ast\cdot q}{q^2}q^\mu +
  (m_B+m_V)A_1(q^2)\left(\epsilon_2^{\ast\mu}-
  \frac{\epsilon_2^\ast\cdot q}{q^2}q^\mu\right)
\nonumber\\
&& -A_2(q^2)\frac{\epsilon_2^\ast\cdot q}{m_B+m_V}
 \left(P_1^\mu+P_2^{\mu} -\frac{m_B^2-m_V^2}{q^2}q^\mu\right),
\end{eqnarray}
with the momentum $q=P_1-P_2$. The form factors $V$, $A_1$,
and $A_2$ satisfy the symmetry relations in the large-energy
limit,
\begin{eqnarray}
&&\frac{m_B}{m_B+m_V} V(q^2) = \frac{m_B+m_V}{2 E} A_1(q^2)
= \xi_\perp(E)\;,
\label{rho1}\\
& &\frac{m_B+m_V}{2 E}A_1(q^2) - \frac{m_B-m_V}{m_B}A_2(q^2)
=\frac{m_V}{E} \xi_\parallel(E)\;, \label{rho2}
\end{eqnarray}
where $E$ is the $V$ meson energy, and the function
$\xi_\parallel(E)$ ($\xi_\perp(E)$) is associated with the
transition into a longitudinally (transversely) polarized $V$
meson. Our definition of $\xi_\parallel$ differs from those in
\cite{BF,Ch}, such that we have $\xi_\perp\approx \xi_\parallel$,
when not distinguishing the longitudinal and transverse
polarizations.


Consider the $B^+\to \rho^{+}\rho^0$ mode as an example, which is
dominated by a tree contribution (assuming that the electroweak
penguin contribution is negligible). The explicit expression of
the relevant effective weak Hamiltonian will not be shown here. In
FA, the decay amplitude is written as
\begin{eqnarray}
\langle \rho^+ \rho^0 | {\cal H}_{\rm eff}|B^+ \rangle &=&
\displaystyle{G_F \over \sqrt{2}}V_{ub}^*V_{ud} \left[a_1 \langle
\rho^+(P_3,\epsilon_3^*)|{\bar u}\gamma_\mu(1-\gamma_5) d|0
\rangle\langle \rho^0(P_2,\epsilon_2^*)|{\bar
b}\gamma^\mu(1-\gamma_5) u|B^+(P_1)
\rangle\right.\nonumber\\
& &\left.+ a_2 \langle \rho^0(P_2,\epsilon_2^*)|{\bar
u}\gamma_\mu(1-\gamma_5) u|0 \rangle\langle
\rho^+(P_3,\epsilon_3^*)|{\bar b}\gamma^\mu(1-\gamma_5) d|B^+(P_1)
\rangle\right]\;,\label{amr}
\end{eqnarray}
where the two terms can be combined, leading to the helicity
amplitudes,
\begin{eqnarray}
A_{L}&\propto&
(1+r_2)\epsilon_2^*(L)\cdot
\epsilon_3^*(L)\left[A_1 -\frac{2\epsilon_2^*\cdot P_3P_2\cdot
\epsilon^*_3}{m_B^2(1+r_2)^2\epsilon_2^*\cdot
\epsilon_3^*}A_2\right]\;, \nonumber\\
A_{\parallel}&\propto&-\sqrt{2}
(1+r_2)A_1\;, \nonumber \\
A_{\perp}&\propto&-
\frac{2r_2r_3}{1+r_2}\sqrt{2[(v_2\cdot v_3)^{2}-1]} V\;.
\label{ase1r}
\end{eqnarray}
The relative phases among the helicity amplitudes are
$\phi_\parallel=\phi_\perp=\pi$ under FA in our convention. In the
large-energy limit, i.e., employing Eqs.~(\ref{rho1}) and
(\ref{rho2}), Eq.~(\ref{ase1r}) becomes
\begin{eqnarray}
A_{L}&\propto& 2
r_2\epsilon_2^*(L)\cdot\epsilon_3^*(L) \xi_\parallel\;, \nonumber\\
A_{\parallel}&\propto&
-2\sqrt{2}
r_2\eta\,\xi_\perp\;,\nonumber\\
A_{\perp}&\propto&- 2
r_2r_3\sqrt{2[(v_2\cdot v_3)^{2}-1]}\, \xi_\perp\;, \label{ase2r}
\end{eqnarray}
implying that QCD dynamics has been absorbed into the two
functions $\xi_\parallel$ and $\xi_\perp$. For a rough estimate,
adopting $\xi_\parallel\approx (f_\rho/f_\rho^T)\xi_\perp$,
$f_\rho=200$ MeV and $f_\rho^T=160$ MeV \cite{BBKT}, the $B^+\to
\rho^{+}\rho^0$ polarization fractions are given by,
\begin{eqnarray}
R_L\sim 0.95\;,\;\;\;\; R_{\parallel}\sim 0.03\;,\;\;\;\;
R_{\perp}\sim 0.02\;, \label{ase3r}
\end{eqnarray}
consistent with the data in Table~\ref{tab:tab2}. The polarization
fractions of other tree-dominated $B\to VV$ modes, including
$\rho^+\rho^-$ and $\rho^+\omega$, can be explained in a similar
way.

We generalize Eq.~(\ref{ase2r}) to the $B^+\to (D_s^{*+},
D^{*+})\rho^0$ modes, which are mainly governed by the $B\to\rho$
form factors, with
the masses $m_{D_s^{*}}$, $m_{D^{*}}$ being substituted for
$m_\rho$, respectively. Their polarization fractions are predicted
to be,
\begin{eqnarray}
& &D_s^{*+}\rho^0:\;\;\; R_L\sim 0.70\;,\;\;\;\;
R_{\parallel}\sim 0.16\;,\;\;\;\;R_{\perp}\sim 0.14\;,\nonumber\\
& &D^{*+}\rho^0:\;\;\; R_L\sim 0.72\;,\;\;\;\; R_{\parallel}\sim
0.15\;,\;\;\;\;R_{\perp}\sim 0.13\;, \label{asr3}
\end{eqnarray}
which can be compared with the data in the future.

\section{SUBLEADING CORRECTIONS}

Away from the heavy quark limit, the form factors in
Eq.~(\ref{bds}) deviate from the IW function. Beyond FA,
nonfactorizable contributions appear. Both corrections, being
subleading \cite{TLS2,KKLL}, can be calculated more reliably in
the PQCD approach based on $k_T$ factorization theorem
\cite{BS,LS} due to the stronger end-point suppression in the
former \cite{TLS2} and to the soft cancellation between a pair of
nonfactorizable diagrams in the latter \cite{CKL}. In this section
we shall examine whether the simple estimates made in the previous
section are robust under these subleading corrections. The $k_T$
factorization theorem for the $B\to D^{*}$ form factors in the
large-recoil region of the $D^{*}$ meson can be proved following
the procedure in \cite{NL}, which are expressed as the convolution
of hard kernels with the $B$ and $D^{*}$ meson wave functions in
both the momentum fractions $x$ and the transverse momenta $k_T$
of partons. A hard kernel, being infrared-finite, is calculable in
perturbation theory. The $B$ and $D^{*}$ meson wave functions,
collecting the soft dynamics in the decays, are not calculable but
universal. After including the parton $k_T$, the end-point
singularities, which usually break QCDF at subleading level, do
not appear, and PQCD factorization formulas are well-defined. This
formalism has been applied to the semileptonic decay $B\to
D^{(*)}l\nu$ \cite{TLS2}, and the nonleptonic decays $B\to
D^{(*)}\pi(\rho)$ \cite{KKLL,WYL} and $B\to D_s^{(*)}D_s^{(*)}$
\cite{LLX04} successfully.

\subsection{$B^0\to D_s^{*+} D^{*-}$}

\begin{figure}[t]
\centerline{
\includegraphics[width=15cm]{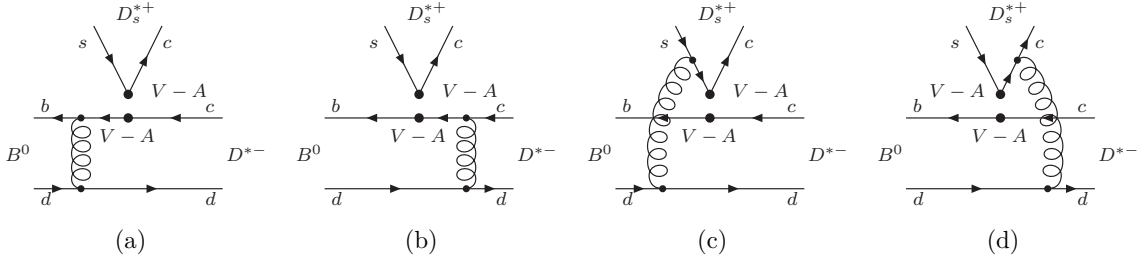}
} \caption{Lowest-order diagrams for the $B^0\to D_s^{*+}D^{*-}$
decay.} \label{fig1}
\end{figure}

For the $B^0\to D_s^{*+} D^{*-}$ mode, we shall neglect the
penguin contribution, which is suppressed by the Wilson
coefficients $C_{3-6}\sim 0.05 a_1$. According to the lowest-order
external-$W$ emission diagrams in Fig.~\ref{fig1}, the decay
amplitudes for different final helicity states are expressed as
\begin{eqnarray}
{\cal
M^{(\sigma)}}&=&\frac{G_F}{\sqrt{2}}V^*_{cb}V_{cs}\left[m_B^2\,
(f_{D_s^*}\, {\cal F}_{L}+{\cal M}_{L}),\; m_B^2
\epsilon^{*}_{2}(T)\cdot\epsilon^{*}_{3}(T)\, (f_{D_s^*}\, {\cal
F}_{N}+{\cal M}_{N}),\right.
\nonumber\\
& &\left. i\, \epsilon^{\alpha \beta\gamma \rho}
\epsilon^{*}_{2\alpha}\epsilon^{*}_{3\beta} P_{2\gamma }P_{3\rho
}\, (f_{D_s^*}\, {\cal F}_{T}+{\cal M}_{T})\right]\;, \label{M1}
\end{eqnarray}
where ${\cal F}_{L,N,T}$ come from the factorizable diagrams,
Figs.~\ref{fig1}(a) and \ref{fig1}(b), and ${\cal M}_{L,N,T}$ from
the nonfactorizable diagrams, Figs.~\ref{fig1}(c) and
\ref{fig1}(d).

We first compute each factorizable amplitude in terms of the
``form factors'' as an expansion in $r_2$,
\begin{eqnarray}
\xi_i&=&\xi
+\xi_i^{({\rm NL})}\;,\;\;i=A_1, A_3, V\;,\nonumber\\
\xi_{A2}&=&\xi_{A2}^{({\rm NL})}\;, \label{xi_form}
\end{eqnarray}
where the superscript NL denotes the next-to-leading-power
corrections. Equation (\ref{xi_form}) is equivalent to the
heavy-quark expansion of the heavy-heavy currents in $1/m_b$ and
in $1/m_c$ \cite{N1}, $m_b$ ($m_c$) being the $b$ ($c$) quark
mass. We refer the explicit expressions of the PQCD factorization
formulas for $\xi$ and $\xi_i^{({\rm NL})}$ to \cite{TLS2}, which
have absorbed the Wilson coefficient $a_1$ in the current
analysis.
In terms of Eq.~(\ref{xi_form}), the factorizable amplitudes are
written as
\begin{eqnarray}
{\cal F}_{L}
  &=&
    \sqrt{r_2}r_3\,
    \left\{
     \epsilon_2^*(L)\cdot \epsilon_3^*(L)(\eta+1)\,
     \left( \xi + \xi_{A1}^{({\rm NL})}\right)
    \right.\nonumber\\
    & &\left.\ \ \ \ \ \ \ \ \
    - \epsilon_2^*(L)\cdot v_1 \,
       \left[ \epsilon_3^*(L)\cdot v_1 \xi_{A2}^{({\rm NL})}
        + \epsilon_3^*(L)\cdot v_2
          \left( \xi + \xi_{A3}^{({\rm NL})}\right)
      \right]
    \right\}\;,
\nonumber\\
{\cal F}_{N}
  &=&
    \sqrt{r_2}r_3\, (\eta+1) \,
    \left( \xi + \xi_{A1}^{({\rm NL})}\right)\;,
\nonumber\\
{\cal F}_{T}
  &=&
    \frac{r_3}{\sqrt{r_2}}\,
    \left( \xi + \xi_V^{({\rm NL})}\right)\;.
\label{xi_sum}
\end{eqnarray}
A numerical analysis gives
\begin{eqnarray}
\xi_{A_1}^{\rm (NL)}\sim -0.02\, \xi\;,\;\;\;\; \xi_{A_2}^{\rm
(NL)}\sim -0.19\, \xi\;,\;\;\;\; \xi_{A_3,V}^{\rm (NL)}\sim
-0.05\, \xi\;. \label{ff1}
\end{eqnarray}
Inserting the deviation from the IW function into
Eq.~(\ref{ase1}), the polarization fractions are modified only
slightly:
\begin{eqnarray}
R_L\sim 0.54\;,\;\;\;\; R_{\parallel}\sim 0.41\;,\;\;\;\;
R_{\perp}\sim 0.05\;. \label{ard1}
\end{eqnarray}
The largest next-to-leading-power correction comes from
$\xi_{A_2}^{\rm (NL)}$, which is, however, suppressed by the
factor $r_2$.

Next we calculate the subleading corrections from the
nonfactorizable diagrams in Figs.~\ref{fig1}(c) and \ref{fig1}(d).
For simplicity, we shall expand all the kinematical factors and
the polarization amplitudes up to next-to-leading power in
$r_{2,3}$.
The factorization formulas are given by
\begin{eqnarray}
{\cal M}_{L}
  &=&
    16\pi C_F \sqrt{2N_c}\,m_B^2
    \int_0^1 dx_1dx_2dx_3
    \int_0^\infty b_1db_1\, b_3db_3\,
    \phi_B(x_1,b_1)\, \phi_{D^*}(x_2)\, \phi_{D_s^*}(x_3)
  \nonumber\\
  & &\times\,
    \left[\left(x_3-x_1+r_2x_2\right)\,
        E_b(t_b^{(1)})\, h_b^{(1)}(x_1,x_2,x_3,b_1,b_3)
  \right.\nonumber \\
  & & \left. \ \ \ \ \ \ \
        - \left( 1+x_2-x_3-x_1-x_2r_2 \right)\,
        E_b(t_b^{(2)})\, h_b^{(2)}(x_1,x_2,x_3,b_1,b_3)
        \right]
\;, \label{mL}
\\
{\cal M}_{N}
  &=&
    16\pi C_F \sqrt{2N_c}\,m_B^2
    \int_0^1 dx_1dx_2dx_3
    \int_0^\infty b_1db_1\, b_3db_3\,
    \phi_B(x_1,b_1)\, \phi_{D^*}(x_2)\, \phi_{D_s^*}(x_3)
  \nonumber\\
  & &\times\,
    \left[ x_3\,r_3\,
        E_b(t_b^{(1)})\, h_b^{(1)}(x_1,x_2,x_3,b_1,b_3)
  \right.\nonumber \\
  & & \left. \ \ \ \ \ \ \
        - \left( 1+x_3\right)\,r_3\,
        E_b(t_b^{(2)})\, h_b^{(2)}(x_1,x_2,x_3,b_1,b_3)
        \right]
\;, \label{mN}
\\
{\cal M}_{T}
  &=&
  2{\cal M}_{N}
\;. \label{mT}
\end{eqnarray}

Radiative corrections to the meson wave functions generate the
double logarithms $\alpha_s\ln^2 k_T$ from the overlap of
collinear and soft enhancements, whose Sudakov resummation has
been studied in \cite{LY1}. The Sudakov factors from $k_T$
resummation for the $B$ meson, the $D^*$ meson and the $D_s^*$
meson are given, according to \cite{LL04}, by
\begin{eqnarray}
\exp[-S_{B}(\mu)]&=&\exp\left[-s(k_1^-,b_1)
-\frac{5}{3}\int_{1/b_1}^\mu
\frac{d{\bar\mu}}{\bar\mu}\gamma(\alpha_s({\bar\mu}))\right]\;,
\nonumber\\
\exp[-S_{D^*}(\mu)]&=&\exp\left[-s(k_2^+,b_2)
-\frac{5}{3}\int_{1/b_2}^\mu
\frac{d{\bar\mu}}{\bar\mu}\gamma(\alpha_s({\bar\mu}))\right]\;,
\nonumber\\
\exp[-S_{D_s^*}(\mu)]&=&\exp\left[-s(k_3^-,b_3)
-\frac{5}{3}\int_{1/b_3}^\mu
\frac{d{\bar\mu}}{\bar\mu}\gamma(\alpha_s({\bar\mu}))\right]\;,
\label{ktd}
\end{eqnarray}
respectively, with the quark anomalous dimension
$\gamma=-\alpha_s/\pi$. The momenta $k_1^-=x_1P_1^-$,
$k_2^+=x_2P_2^+$, and $k_3^-=x_3P_3^-$, carried by the light
valence quarks in the $B$, $D^*$, and $D_s^*$ mesons,
respectively, define the momentum fractions $x$. The impact
parameters $b_1$, $b_2$, and $b_3$ are conjugate to the transverse
momenta carried by the light valence quarks in the $B$, $D^*$, and
$D_s^*$ mesons, respectively. For the explicit expression of the
Sudakov exponent $s$, refer to \cite{KLS}. Note that the
coefficient $5/3$ of the quark anomalous dimension in
Eq.~(\ref{ktd}) differs from 2 for a light meson. The reason is
that the rescaled heavy-quark field adopted in the definition of a
heavy-meson wave function has a self-energy correction different
from that of the full heavy-quark field \cite{LL04}. The evolution
factors are then given by
\begin{eqnarray}
E_b(t) &=& \alpha_s(t)\frac{C_1(t)}{N_c}
           \exp\left[-S(t)|_{b_2=b_1}\right]\;,
\end{eqnarray}
with the Sudakov exponent $S=S_B+S_{D^*}+S_{D^*_s}$.

The functions $h^{(j)}_b$, $j=1$ and $2$, are written as
\begin{eqnarray}
h^{(j)}_b
  &=&
    \left[\theta(b_1-b_3)
          K_0\left(Bm_B b_1\right)
          I_0\left(Bm_Bb_3\right)
    \right. \nonumber \\
  & &\quad \left.
    + \theta(b_3-b_1)
       K_0\left(Bm_B b_3\right)
       I_0\left(Bm_B b_1\right)
    \right]
  \nonumber \\[1mm]
  & & \times
    \left(
    \begin{array}{cc}
      K_{0}\left(B_{j}m_Bb_{3}\right) &  \mbox{for $B^2_{j} \geq 0$}
      \\[1mm]
      \frac{i\pi}{2} H_{0}^{(1)}\left(\sqrt{|B_{j}^2|}m_Bb_{3}\right) &
      \mbox{for $B^2_{j} \leq 0$}
    \end{array} \right)\;,
\end{eqnarray}
with the variables
\begin{eqnarray}
B^{2} &=& x_1 x_2 r_2\eta^+\;,
\nonumber \\
B_{1}^{2} &=& x_1 x_2 r_2\eta^+
          - x_2 x_3 \left(r_2\eta^+ - r_2^2\right)\;,
\nonumber \\
B_{2}^{2} &=& x_1 x_2r_2\eta^+
          - x_2 (1-x_3)\left(r_2\eta^+-r_2^2\right)
          + (x_1+x_3)\left(1-r_2\eta^+\right)
          + x_3\left(r_2^2-r_2\eta^-\right)\;.
\label{mis}
\end{eqnarray}
The scales $t_b^{(j)}$ are chosen as
\begin{eqnarray}
t_b^{(j)} &=&
  {\rm max}\left(Bm_B,\sqrt{|B_j^2|}m_B,1/b_1,1/b_3\right)\;.
\end{eqnarray}

The $B$ and $D^*$ meson wave functions involved in
Eqs.~(\ref{mL})-(\ref{mT}) are also referred to \cite{TLS2}:
\begin{eqnarray}
\phi_B(x,b)&=&N_Bx^2(1-x)^2
\exp\left[-\frac{1}{2}\left(\frac{xm_B}{\omega_B}\right)^2
-\frac{\omega_B^2 b^2}{2}\right]\;, \label{os}\\
\phi_{D^*}(x)&=&\frac{3f_{D^*}}{\sqrt{2N_c}}x(1-x)
\left[1+C_{D^*}(1-2x)\right]\;.
\end{eqnarray}
The $D_s^*$ meson wave function $\phi_{D_s^*}$ will be assumed to
have the same functional form as $\phi_{D^*}$. We do not consider
the $b$ dependence of $\phi_{D^*}$ in the large recoil region of
the $D^*$ meson \cite{TLS2}. The Gaussian form of $\phi_B$ was
motivated by the oscillator model in \cite{BW}. The shape
parameter $\omega_B=0.40$ GeV comes from \cite{TLS}, and
$C_{D^*}=1.04$ is determined from the IW function $\xi(\eta=1.3)=
0.7$. The normalization constant $N_B$ is related to the decay
constant $f_B$ through
\begin{eqnarray}
\int dx\phi_B(x,b=0)=\frac{f_B}{2\sqrt{2N_c}}\;.
\end{eqnarray}
We shall not distinguish the $D^*$ meson wave functions associated
with the longitudinal and transverse polarizations. There are
various models of the $B$ meson wave functions available in the
literature \cite{LL04}. It has been confirmed that the model in
Eq.~(\ref{os}) and the model derived in \cite{KQT} with a
different functional form lead to similar numerical results for
the $B\to\pi$ form factor \cite{WY}.

Taking into account only the nonfactorizable corrections, the
polarization fractions become
\begin{eqnarray}
R_L\sim 0.54\;,\;\;\;\; R_{\parallel}\sim 0.40\;,\;\;\;\;
R_{\perp}\sim 0.06\;. \label{ardn1}
\end{eqnarray}
Including both the subleading factorizable and nonfactorizable
corrections, we have
\begin{eqnarray}
R_L\sim 0.56\;,\;\;\;\; R_{\parallel}\sim 0.39\;,\;\;\;\;
R_{\perp}\sim 0.06\;, \label{ardt1}
\end{eqnarray}
which are still close to the values in Eq.~(\ref{arr}). Hence, the
simple kinematic estimate made in the heavy-quark limit is very
reliable.

At last, we compute the $B^0\to D_s^{*+}D^{*-}$ branching ratio,
considering only the leading contribution. Employing the CKM
matrix elements $V_{cb}\, =\, 0.0412$, $V_{cs}\, =\, 0.996$, and
$V_{cd}\, =\, -0.224$, the
quark masses $m_b\, =\, 4.8 \; {\rm GeV}$ and $m_t\, =\, 174.3 \;
{\rm GeV}$, the meson decay constants $f_{B}\, =\, 200\; {\rm
MeV}$, $f_{D^*}\,=\, 230\; {\rm MeV}$, and $f_{D_s^*}\, =\, 240
\;{\rm MeV}$,
the lifetimes $\tau_{B^0}\, =\, 1.542\times 10^{-12}\;{\rm sec}$
and $\tau_{B^\pm}\, =\, 1.674\times 10^{-12} \;{\rm sec}$, and the
Fermi constant $G_F\, =\, 1.16639\times 10^{-5}\;{\rm GeV}^{-2}$,
we predict
\begin{eqnarray}
B(B^0\to D_s^{*+} D^{*-})= \left(2.6^{+1.1}_{-0.8}\right) \%
\;,\label{brds}
\end{eqnarray}
which is consistent with the recent measurement $(1.85\pm 0.09\pm
0.16)\%$ \cite{Bar040}. The theoretical uncertainty in
Eq.~(\ref{brds}) arises from the allowed range of the shape
parameter $\omega_B=(0.40\pm 0.04)$ GeV \cite{TLS} in the $B$ meson
wave function. Note that the polarization fractions are
insensitive to this overall source of uncertainty.

\subsection{$B\to D^*D^*$}

\begin{figure}[t]
\centerline{
\includegraphics[width=15cm]{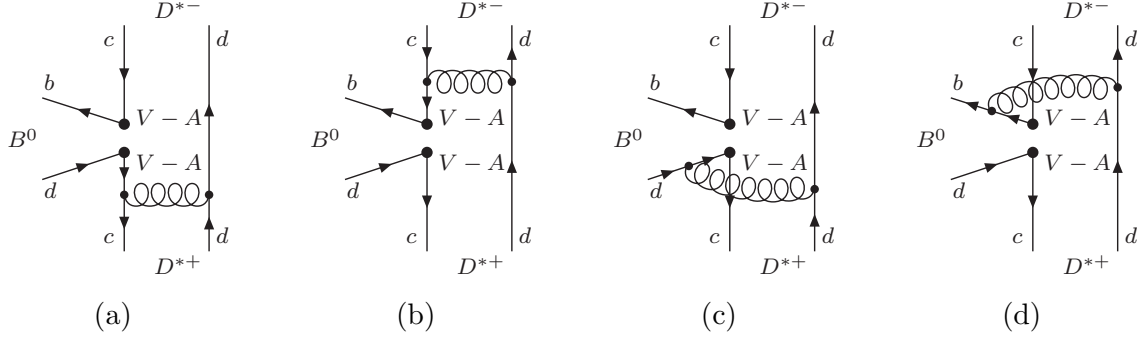}
} \caption{$W$-exchange diagrams for the $B^0\to D^{*+}D^{*-}$
decay.} \label{fig_dpdm}
\end{figure}

For the $B^0\to D^{*+}D^{*-}$ mode, there exists an additional
contribution from the $W$-exchange topology shown in
Fig.~\ref{fig_dpdm} compared to $B^0\to D_s^{*+}D^{*-}$. The
factorizable $W$-exchange diagrams, Figs.~\ref{fig_dpdm}(a) and
\ref{fig_dpdm}(b), vanish exactly because of the helicity
suppression. An explicit evaluation shows that the nonfactorizable
$W$-exchange diagrams, Figs.~\ref{fig_dpdm}(c) and
\ref{fig_dpdm}(d), contribute only 2\% of the external-$W$
emission ones in Fig.~\ref{fig1}. Therefore, the $W$-exchange
topology is even less important than the penguin one. Note that
the $W$-exchange contribution is as important as the
nonfactorizable contribution for the $B\to D^{(*)}\pi(\rho)$
decays \cite{KKLL}, to which the helicity suppression does not
apply. The $B^0\to D^{*+}D^{*-}$ factorization formulas are then
similar to those of $B^0\to D_s^{*+}D^{*-}$ but with the
appropriate replacements of the $D_s^*$ meson mass, decay
constant, and wave function by the $D^*$ meson ones, respectively.
Similarly, if neglecting the $W$-exchange and penguin
contributions, the $B^+\to D^{*+}{\bar D}^{*0}$ decay amplitudes
will be the same as of $B^0\to D^{*+}D^{*-}$.

The numerical results are summarized below. Including both the
subleading factorizable and nonfactorizable corrections, we obtain
the polarization fractions of the $B^0\to D^{*+}D^{*-}$ and
$B^+\to D^{*+}{\bar D}^{*0}$ modes,
\begin{eqnarray}
R_L\sim 0.56\;,\;\;\;\;
R_{\parallel}\sim 0.38\;,\;\;\;\;
R_{\perp}\sim 0.06\;,
\label{ard2}
\end{eqnarray}
which are also close to the simple estimate in Eq.~(\ref{ard}).
Using the same input parameters for the quark masses, the $B$
meson lifetimes,..., we predict the branching ratio,
\begin{eqnarray}
B(B^0\to D^{*+} D^{*-})= \left(1.2\,{}^{+0.5}_{-0.3}\right)\times
10^{-3} \;,\label{bdd}
\end{eqnarray}
which is consistent with the updated measurement $B(B^0\to D^{*+}
D^{*-})=(8.1\pm 0.8\pm 0.1)\times 10^{-4}$ \cite{Belich}. The
branching ratio $B(B^+\to D^{*+} D^{*0})$ can be obtained simply
by changing the $B$ meson lifetime, whose value is close to that
in Eq.~(\ref{bdd}).

\begin{figure}[t]
\centerline{
\includegraphics[width=15cm]{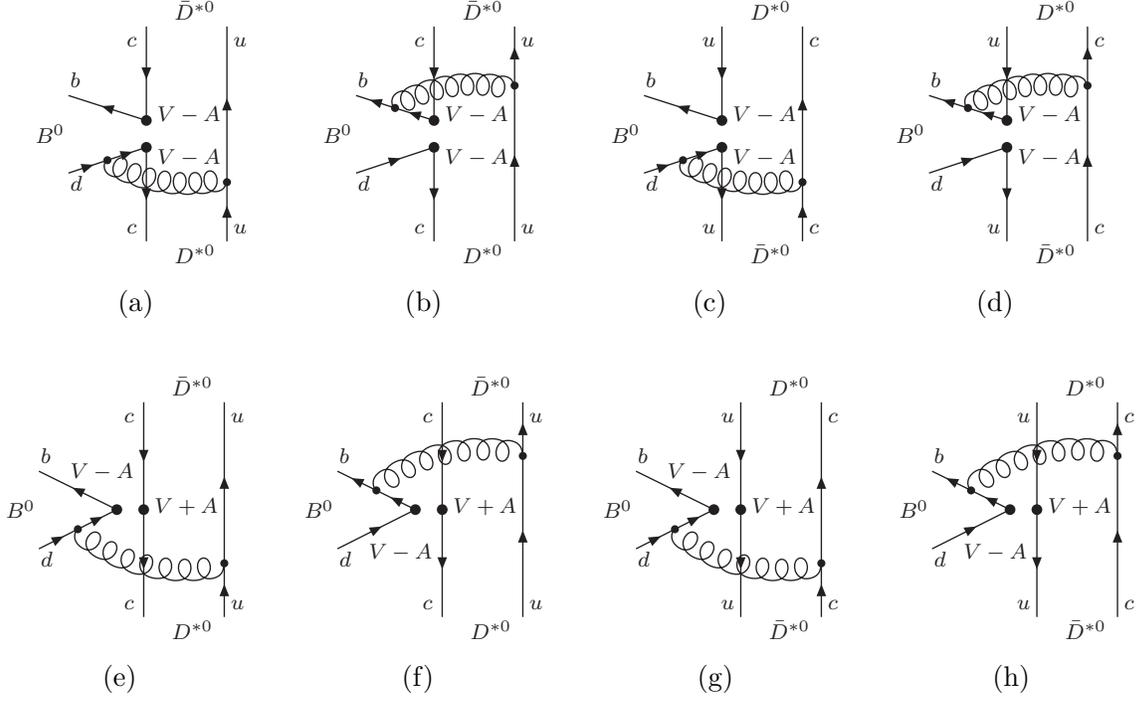}
} \caption{Lowest-order diagrams for the $B^0\to \bar
D^{*0}D^{*0}$
 decay.}
\label{fig_d0d0}
\end{figure}

Since only the $W$-exchange topology in
Figs.~\ref{fig_d0d0}(a)-(d) and the penguin annihilation topology
in Figs.~\ref{fig_d0d0}(e)-(h) contribute, the $B^0\to {\bar
D}^{*0}D^{*0}$ decay must have a tiny branching ratio. For both
topologies, the factorizable amplitudes diminish because of the
helicity suppression (only the penguin annihilation from the
$(S-P)(S+P)$ operators survives, which do not exist in this mode).
Hence, they are not exhibited in Fig.~\ref{fig_d0d0}. With the
penguin amplitudes being down by the Wilson coefficients, we shall
calculate only the tree contribution. By measuring the $B^0\to
{\bar D}^{*0}D^{*0}$ mode, we learn how the nonfactorizable
$W$-exchange contribution governs the polarization fractions. It
will be shown that the polarization fractions in this decay differ
dramatically from those in the $B^0 \to D^{*+}D^{*-}$ and $B^+ \to
D^{*+}\bar D^{*0}$ decays. On the other hand, it has been proposed
\cite{ADL04} to extract the weak phase $\phi_3$ from the $B\to
D^{(*)}D^{(*)}$, $D_s^{(*)}D^{(*)}$ measurements.

The amplitudes from Figs.~\ref{fig_d0d0}(a) and \ref{fig_d0d0}(b)
are given by
\begin{eqnarray}
{\cal M}_{L}
  &=&
    16\pi C_F \sqrt{2N_c}\,m_B^2
    \int_0^1 dx_1dx_2dx_3
    \int_0^\infty b_1db_1\, b_2db_2\,
    \phi_B(x_1,b_1)\, \phi_{D^*}(x_2)\, \phi_{D^*}(x_3)
  \nonumber\\
  & &\times\,
    \left[ \left( x_3-x_1 \right)\,
        E_f(t_f^{(1)})\, h_f^{(1)}(x_1,x_2,x_3,b_1,b_2)
  \right.\nonumber \\
  & & \left. \ \ \ \ \ \ \
        -x_2\,
        E_f(t_f^{(2)})\, h_f^{(2)}(x_1,x_2,x_3,b_1,b_2)
        \right]\;, \label{excmL}
\\
{\cal M}_{N,T} &=& O\left(r_i^2\right) \;, \label{excmNT}
\end{eqnarray}
and those from Figs.~\ref{fig_d0d0}(c) and \ref{fig_d0d0}(d) are
\begin{eqnarray}
{\cal M}'_L
  &=&
    16\pi C_F \sqrt{2N_c}\,m_B^2
    \int_0^1 dx_1dx_2dx_3
    \int_0^\infty b_1db_1\, b_2db_2\,
    \phi_B(x_1,b_1)\, \phi_{D^*}(x_2)\, \phi_{D^*}(x_3)
  \nonumber\\
  & &\times\,
    \left[ \left( 1-x_2 \right) \,
        E_f(t_{f}^{'(1)})\, h_{f}^{'(1)}(x_1,x_2,x_3,b_1,b_2)
  \right.\nonumber \\
  & & \left. \ \ \ \ \ \ \
        - \left( 1- x_3 + x_1 \right)\,
        E_f(t_{f}^{'(2)})\, h_{f}^{'(2)}(x_1,x_2,x_3,b_1,b_2)
        \right]
\;, \label{excmL2}
\\
{\cal M}'_{N,T} &=& O\left(r_i^2\right) \;. \label{excmNT2}
\end{eqnarray}
It is observed that the transverse polarization amplitudes vanish
in the current next-to-leading-power accuracy. The evolution
factors are
\begin{eqnarray}
E_f(t) &=& \alpha_s(t)\frac{C_2(t)}{N_c}
           \exp\left[-S(t)|_{b_3=b_2}\right]\;.
\end{eqnarray}
The hard functions $h^{(\prime)(j)}_{f}$, $j=1$ and $2$, are
written as
\begin{eqnarray}
h^{(\prime)(j)}_{f}
  &=&
    \frac{i\pi}{2}
    \left[\theta(b_1-b_2)
       H_0^{(1)}\left(F^{(\prime)}m_B b_1\right)
       J_0\left(F^{(\prime)}m_Bb_2\right)
    \right. \nonumber \\
  & &\quad \left.
    + \theta(b_2-b_1)
       H_0^{(1)}\left(F^{(\prime)}m_B b_2\right)
       J_0\left(F^{(\prime)}m_B b_1\right)
    \right]
  \nonumber \\[1mm]
  & & \times
    \left(
    \begin{array}{cc}
      K_{0}\left(F^{(\prime)}_{j}m_Bb_{1}\right) &
      \mbox{for $F^{(\prime)2}_{j} \geq 0$} \\[1mm]
      \frac{i\pi}{2} H_{0}^{(1)}
      \left(\sqrt{|F^{(\prime)2}_{j}|}m_Bb_{1}\right) &
      \mbox{for $F^{(\prime)2}_{j} \leq 0$}
    \end{array} \right)\;,
\end{eqnarray}
with the variables
\begin{eqnarray}
F^{2} &=& x_2 r_2\eta^+ x_3(1-r_2\eta^-)\;,
\nonumber \\
F_{1}^{2} &=& x_2 r_2\eta^+
              \left[ x_1 - x_3(1-r_2\eta^-) \right]\;,
\nonumber \\
F_{2}^{2} &=& 1+(1-x_2r_2\eta^+)
              \left[ x_1 - 1 + x_3(1-r_2\eta^-) \right]\;,
\nonumber \\
F^{\prime\, 2} &=&
          (1-x_2r_2\eta^+)
          \left[ 1 - x_3(1-r_2\eta^-) \right]\;,
\nonumber \\
F_{1}^{\prime\, 2} &=&
           (1-x_2r_2\eta^+)
            \left[ x_1 -1 +  x_3(1-r_2\eta^-) \right]\;,
\nonumber \\
F_{2}^{\prime\, 2} &=&
             1+ x_2r_2\eta^+
              \left[ x_1 - x_3(1-r_2\eta^-) \right]
\;.
\end{eqnarray}
The variables $F^\prime$'s can be obtained from $F$'s by
interchanging $x_2r_2\eta^+$ and $1-x_2r_2\eta^+$, and
$x_3(1-r_2\eta^-)$ and $1 - x_3(1-r_2\eta^-)$.
The hard scales $t_{f}^{(\prime)(j)}$ are chosen as
\begin{eqnarray}
t_{f}^{(\prime)(j)}
  &=& {\rm max}
    \left(F^{(\prime)}m_B,\sqrt{|F^{(\prime)2}_j|}m_B,1/b_1,1/b_2\right)\;.
\end{eqnarray}

The longitudinal polarization fraction dominates:
\begin{eqnarray}
R_L\sim 1\;,\;\;\;\; R_{\parallel}\sim R_{\perp}\sim {\rm
few}\%\;, \label{ard3}
\end{eqnarray}
which differs very much from that in Eq.~(\ref{ard2}). Therefore,
the comparison of the theoretical prediction with the future data
can test the PQCD approach. We also predict the branching ratio,
\begin{eqnarray}
B(B^0\to {\bar D}^{*0}D^{*0}) =
\left(8.9^{+1.4}_{-1.1}\right)\times 10^{-5} \;,
\end{eqnarray}
which can also be compared with the future data.

\subsection{$B\to\rho\rho$}

In this subsection we examine whether the simple estimate of the
polarization fractions in the tree-dominated $B$ meson decays into
two light vector mesons is robust under subleading corrections.
Similar to the previous subsection, the $O(m_V/m_B)$ terms should
be included into the factorizable amplitudes at this level of
accuracy. At the same time, the two-parton twist-4 contribution
appears, since the linear end-point singularity involved in
collinear factorization theorem modifies the power behavior from
$O(m_V^2/m_B^2)$ into $O(m_V/m_B)$ \cite{TLS}. The inclusion of
these two corrections makes complete the next-to-leading-power
analysis at the two-parton level. The nonfactorizable amplitudes
have been known to be small due to the strong cancellation between
a pair of nonfactorizable diagrams \cite{KLS,LUY}. The
annihilation amplitudes for tree-dominated modes are also
negligible due to helicity suppression \cite{CKL}. Hence, we shall
not consider the two-parton twist-4 correction to these two
subleading contributions. Below we analyze the $B\to\rho\rho$
longitudinal polarization amplitude as an example.

The two-parton $\rho$ meson distribution amplitudes up to twist 4
are defined by the following expansion \cite{BBKT},
\begin{eqnarray}
\langle \rho^-(P_2,\epsilon_{2L}^*)|\bar d(z)_ju(0)_l|0\rangle
&=&\frac{1}{\sqrt{2N_c}}\int_0^1 dx e^{ixp_2\cdot z}\Bigg[\not
p_2\phi_{\rho}(x) +m_{\rho}(\not n_+\not n_--1)
\phi_{\rho}^{t}(x)\nonumber\\
& &\;\;\;\;\;\;+m_{\rho}
I\phi_{\rho}^s(x)-\frac{m_{\rho}^2}{2p_2\cdot n_-}\not
n_-\phi_{\rho}^{g}(x)\Bigg]_{lj}\;, \label{klpf}
\end{eqnarray}
where the new vector $p_2$ contains only the plus (large)
component of $P_2$. The distribution amplitude $\phi_\rho$ is of
twist 2 (leading twist), $\phi_{\rho}^{t}$ and $\phi_{\rho}^{s}$
of twist 3, and $\phi_\rho^g$ of twist 4. The twist-3 distribution
amplitudes in fact give leading-power contribution due to the
similar modification from the end-point singularity. The explicit
expressions of the above $\rho$ meson distribution amplitudes are
referred to \cite{TLS}, and $\phi_\rho^g$ is given by \cite{BB98}
\begin{eqnarray}
\phi_\rho^g(x)=\frac{f_\rho}{2\sqrt{2N_c}}\left[1-1.62C_2^{1/2}(2x-1)
-0.41C_4^{1/2}(2x-1)\right]\;,
\end{eqnarray}
with the Gegenbauer polynomials,
\begin{eqnarray}
& &C_2^{1/2}(t)=\frac{1}{2}(3t^2-1)\;,\;\;\;
C_4^{1/2}(t)=\frac{1}{8}(35 t^4 -30 t^2 +3)\;.
\end{eqnarray}

The longitudinal factorizable amplitude in the $B\to\rho\rho$
decays is written, up to twist 4, as,
\begin{eqnarray}
{\cal F}_{L} &=& 8  \pi C_F m_B^2 \int_0^1 {\rm d}x_1 {\rm d}x_2
\int_0^{\infty} b_1{\rm d}b_1\, b_2{\rm d}b_2\, \phi_B(x_1,b_1)
\nonumber \\
& &\hspace{5mm}\times\, \bigg\{ \left[ \left( (1+x_2)(1-r_2^2)
-(1+2x_2)r_2^2 \right)\,\phi_{\rho}(x_2)
\right. \nonumber\\
& & \left.\hspace{13mm} + r_2(1-2x_2) \left(
\phi_{\rho}^s(x_2)+\phi_{\rho}^t(x_2) \right) \right]
E_{e}(t^{(1)}_e) h_{e}(x_1,x_2,b_1,b_2)
\nonumber\\
& & + r_2 \left[ 2\, \phi_{\rho}^s(x_2) + r_2\,
\phi_{\rho}^{g}(x_2) \right] E_{e}(t^{(2)}_e)
h_{e}(x_2,x_1,b_2,b_1) \bigg\} \;,
\end{eqnarray}
where the first (second) term containing $E_{e}(t^{(1)}_e)$
$[E_{e}(t^{(2)}_e)]$ comes from the lowest-order diagram similar
to Fig.~\ref{fig1}(a) [Fig.~\ref{fig1}(b)], but with the $D_s^*$
and $D^*$ mesons being replaced by the $\rho$ mesons. The
evolution factor is
\begin{eqnarray}
E_e(t)=\alpha_s(t)\,a_1(t)\,\exp[-S_B(t)-S_{\rho}(t)]
\;,\label{eet}
\end{eqnarray}
with the Sudakov factor from the $k_T$ resummation,
\begin{eqnarray}
\exp[-S_{\rho}(\mu)]=\exp\left[-s(k_2^+,b_2)-s(P_2^+-k_2^+,b_2)
-2\int_{1/b_2}^\mu
\frac{d{\bar\mu}}{\bar\mu}\gamma(\alpha_s({\bar\mu}))\right]\;,
\label{srho}
\end{eqnarray}
and the hard scales,
\begin{eqnarray}
t^{(1)}_e&=&{\rm max}(\sqrt{x_2}m_B,1/b_1,1/b_2)\;, \;\;
t^{(2)}_e={\rm max}(\sqrt{x_1}m_B,1/b_1,1/b_2)\;.
\end{eqnarray}

The hard function is given by
\begin{eqnarray}
h_e(x_1,x_2,b_1,b_2) &=& S_t(x_2)\,
K_{0}\left(\sqrt{x_1x_2}m_Bb_1\right)
\left[\theta(b_1-b_2)K_0\left(\sqrt{x_2}m_B
b_1\right)I_0\left(\sqrt{x_2}m_Bb_2\right)\right.
\nonumber \\
& & \hspace{33mm}
\left.+\theta(b_2-b_1)K_0\left(\sqrt{x_2}m_Bb_2\right)
I_0\left(\sqrt{x_2}m_Bb_1\right)\right]\;.
\end{eqnarray}
The Sudakov factor $S_t(x)$ arises from the threshold resummation
of the double logarithms $\alpha_s\ln^2 x$, which are produced by
the radiative corrections to the hard kernels. Its expression
\cite{UL},
\begin{eqnarray}
S_t(x)=\frac{2^{1+2c}\Gamma(3/2+c)}{\sqrt{\pi}\Gamma(1+c)}
[x(1-x)]^c\;,
\end{eqnarray}
with the constant $c\sim 0.3$, provides further suppression in the
end-point region of $x\to 0$, and improves the perturbative
calculation. Since we have performed the leading-logarithm
resummation so far, only the behavior of $S_t(x)$ at small $x$ is
reliable. The parametrization in the large $x$ region, also
vanishing, was proposed for convenience \cite{UL}.

The numerical results are listed in Table~\ref{t4}, where diagram
(a) [diagram (b)] refers to the lowest-order diagram with the hard
gluon being on the $B$ ($\rho$) meson side. It indicates that the
next-to-leading-power corrections decrease (increase) the
leading-power contribution from diagram (a) [diagram (b)]
slightly. Considering the net effect, these corrections are indeed
negligible.

\begin{table}[tp]
\begin{center}
\begin{tabular}{cccc}
\hline\hline & Diagram (a) & Diagram (b) & sum
\\
\hline
Leading-power & $0.330$ & $0.088$ & $0.418$ \\
Plus next-to-leading power & $0.324$ & $0.096$ & $0.420$ \\
\hline\hline
\end{tabular}
\caption{Contributions to ${\cal F}_{L}$ up to next-to-leading
power. \label{t4}}
\end{center}
\end{table}

\section{COMMENTS ON PLAUSIBLE EXPLANATIONS}

We now comment on the mechanism proposed in the literature to
explain the abnormal polarization fractions of the $B\to \phi K^*$
decays. It will be argued that these proposals involve many free
parameters, can not account for the polarizations of all $B\to VV$
modes simultaneously, or are too small to achieve the purpose.
When discussing the annihilation effect on penguin-dominated
decays, we found that the $B\to\rho K^*$ polarization data can be
understood, which form the third category mentioned in the
Introduction.

\subsection{Annihilation Contribution}

If the power-suppressed annihilation contribution from the
$(S-P)(S+P)$ penguin operators is enhanced by some mechanism, one
could have the different counting rules as shown in
Eq.~(\ref{mod}), and it might be possible to reach $R_L\sim 0.5$.
This is the strategy adopted in \cite{AK}, whose analysis was
performed in the QCDF approach \cite{BBNS}. In QCDF an
annihilation amplitude is not calculable due to the end-point
singularity, and has to be formulated in terms of several free
parameters, such as $\rho_A$. In the current case, different
$\rho_A$ have been introduced for the longitudinal and transverse
polarization amplitudes in order to fit the data. These parameters
greatly reduce the predictive power of QCDF. Because varying free
parameters to explain the data can not be conclusive, we shall
estimate the annihilation contribution in the PQCD approach,
viewing that the PQCD predictions for the penguin annihilation are
consistent with the measured direct CP asymmetries in $B^0\to
K^+\pi^-$, $\pi^+\pi^-$ \cite{KLS,LUY}. Such a calculation for a
pure-penguin $VV$ mode has been performed in \cite{CKL2}. As shown
in Table~\ref{tab5}, both the penguin annihilation and
nonfactorizable contributions help reduce $R_L$. However, the
combined effect is still not sufficient to lower the fractions
$R_L$ of the $B\to\phi K^*$ decays down to around 0.5.

Note that our predicted relative strong phases among $A_{L}$,
$A_{\parallel}$, and $A_{\perp}$ are consistent with the $B\to\phi
K^{*0}$ data:
\begin{eqnarray}
& &\phi_{\parallel}=2.21\pm 0.22\pm 0.05\,
(rad.)\;,\;\;\;\;\phi_{\perp}=2.42\pm 0.21 \pm
0.06\,(rad)\,\cite{Zhang04} \;,\nonumber\\
& &\phi_{\parallel}=2.34^{+0.23}_{-0.20}\pm 0.05\,
(rad.)\;,\;\;\;\;\phi_{\perp}=2.47\pm 0.25\pm
0.05\,(rad)\,\cite{Bar017}\;.
\end{eqnarray}
Table~\ref{tab5} also implies that $R_L$ can decrease down to 0.75
for a pure-penguin $VV$ mode, after taking into account the
penguin annihilation and nonfactorizable contributions. Since the
$B^+\to \rho^+K^{*0}$ decay is a pure-penguin process, and the
$\rho$ meson mass is not very different from the $\phi$ meson
mass, the above PQCD analysis applies. Hence, we expect the
longitudinal fraction $R_L\sim 0.75$ for the $B^+\to \rho^+K^{*0}$
decay, which is consistent with the Babar measurement, but a bit
larger than the Belle measurement. Due to the large uncertainty of
the Belle data, there is in fact no discrepancy.

\begin{table}[htbp]
\begin{center}
\begin{tabular}{cccccc}
\hline\hline
Mode & $ |A_{L}|^{2}$ & $ |A_{\parallel}|^{2}$ & $
|A_{\perp}|^{2}$ & $\phi_{\parallel}(rad.)$ & $\phi_{\perp}(rad.)$
\\ \hline
$\phi K^{*0}$(I) & $0.923$  & $0.040$ & $0.035$ &
$\pi$ & $\pi$\\
\hspace{0.7cm}(II)&  $0.860$ & $0.072$ & $0.063$ & $3.30$
& $3.33$ \\
\hspace{0.7cm}(III) &  $0.833$ & $0.089$ & $0.078$ &
$2.37$ & $2.34$ \\
\hspace{0.7cm}(IV) & { $0.750$} & $0.135$ & $0.115$ & $2.55$ & $2.54$ \\
\hline $\phi K^{*+}$(I)  &  $0.923$ & $0.040$ & $0.035$ &
$\pi$  & $\pi$  \\
\hspace{0.7cm}(II)  &$0.860$ & $0.072$ & $0.063$ & $3.30$ & $3.33$
\\
\hspace{0.7cm}(III) &$0.830$ & $0.094$ & $0.075$ &
$2.37$ & $2.34$ \\
\hspace{0.7cm}(IV)  &{ $0.748$} & $0.133$ & $0.111$ & $2.55$ & $2.54$ \\
\hline\hline
\end{tabular}
\end{center}
\caption{(I) Without nonfactorizable and annihilation
contributions, (II) add  only nonfactorizable contribution, (III)
add only annihilation contribution, (IV) add both nonfactorizable
and annihilation contributions.}\label{tab5}
\end{table}

We then come to another mode $B^+\to \rho^0K^{*+}$, to which the
tree operators contribute. Adopt the Babar measurement $R_L\sim
0.8$ for the $B^+\to \rho^+K^{*0}$ decay, and assume that the tree
contribution, which obeys the counting rules in Eq.~(\ref{nai}),
affects only the longitudinal fraction. Hence, we simply add the
color-allowed and color-suppressed tree amplitudes $T+C\sim 0.6
\exp(-90^oi)P$, which was extracted from the $B\to K\pi$ data
\cite{Charng}, to the $B^+\to \rho^+K^{*0}$ longitudinal
polarization amplitude $P$. The phase $-90^o$ has included the
weak phase $\phi_3\sim 60^o$. Without an explicit computation, we
derive the polarization fractions for the $B^+\to \rho^0K^{*+}$
decay,
\begin{eqnarray}
R_L\sim 0.86\;,\;\;\;\;R_\parallel\sim R_\perp\sim 0.07\;,
\end{eqnarray}
which are consistent with the data within $1\sigma$. We emphasize
that we did not attempt a rigorous calculation of the $B\to \rho
K^*$ decays here, which deserves a separate paper. In the
perturbation theories, such as PQCD and QCDF, the factorizable
$B\to \pi K$ amplitudes and the factorizable $B\to \rho K^*$
amplitudes with longitudinal polarizations are very similar. Small
differences arise only from the meson masses and the distribution
amplitudes. Therefore, the estimation using the three amplitudes
from the $B\to \pi K$ modes, which have been available in the
literature, makes sense.

The analysis and the result of the modes $B\to \omega K^*$ should
be similar to those of $B\to \rho^0 K^*$. The explicit PQCD
analysis of the $B\to\rho (\omega) K^*$ polarizations will be
performed elsewhere. In conclusion, it is not difficult to
accommodate the polarization data of the third category, the $B\to
\rho K^{*}$ decays, within the Standard Model by means of the
penguin annihilation and nonfactorizable contributions. It is also
interesting to propose that the measurement of the $B\to\omega
K^*$ polarizations can test the PQCD approach.

\subsection{Charming Penguin}

A charming penguin arises from the nonperturbative dynamics
involved in a charm quark loop \cite{charming}. It is not
calculable, has to be parameterized as a free parameter, and could
be as large as a leading contribution. Recently, it has been
introduced into soft-collinear effective theory (SCET) in order to
account for the large $B\to\pi^0\pi^0$ branching ratio
\cite{BPRS}. The inputs of the measured CP asymmetries
$S_{\pi\pi}$ and $A_{\pi\pi}$ demand a complex charming penguin,
leading to a large
penguin-over-tree ratio $|P/T|\sim 0.7$ \cite{BPRS}. With this
$P/T$ from the data, a large branching ratio
$B(B^0\to\pi^0\pi^0)\sim 1.9\times 10^{-6}$ was obtained. SCET
does not attempt to explain why $|P/T|$ is so large, even though
the Standard Model calculations based on PQCD and QCDF give
$|P/T|=0.23$-0.29. Another concern is that the $B\to\pi$ form
factor from the data fitting is as small as 0.17 in the presence
of the large charming penguin, in conflict with the values 0.28
from lattice QCD \cite{DB02} and from light-cone sum rules
\cite{KR,PB3}.

It has been also proposed that the charming penguin may be large
enough to modify the counting rules in Eq.~(\ref{nai}), and to
explain the abnormal $B\to\phi K^*$ polarization data \cite{BPRS}.
However, one also requires different free parameters for the
different helicity amplitudes in order to lower the longitudinal
polarization fraction, and to enhance the transverse polarization
fractions. In this sense, SCET is similar to QCDF \cite{AK}, where
different $\rho_A$ were introduced for the different helicity
amplitudes. One needs different parameters for different modes
too, such as $B\to\rho K^*$ and $B\to\phi K^*$. Our comment on
SCET is then the same as on QCDF: the explanation by introducing
as many parameters as necessary is always plausible, but can not
be conclusive. We point out that the current SCET formalism is
only of leading power: the chirally enhanced terms, proportional
to $m_0/m_B$, have been dropped, and the annihilation (or
$W$-exchange) amplitudes have not yet been formulated. We
speculate that if the annihilation amplitude is included into
SCET, the charming penguin may not be so essential. On the other
hand, the charm-loop correction is well-behaved in perturbation
theory without any infrared singularity, which has been known as
the Bander-Silverman-Soni mechanism \cite{BSS}, implying that its
nonperturbative piece is unlikely to be large. Besides, the
light-cone-sum-rule analysis has supported a small charming
penguin \cite{KMM}.

We have taken this chance to investigate the charm-quark loop
correction to the $B\to\phi K^*$ polarization fractions in the
PQCD approach. The gluon invariant mass attaching the charm-quark
loop can be defined unambiguously as
\begin{eqnarray}
q^2=(1-x_2)x_3m_B^2-|{\bf k}_{2T}-{\bf k}_{3T}|^2\;,
\end{eqnarray}
with $x_2$ and $k_{2T}$ ($x_3$ and $k_{3T}$) being the momentum
fraction and the transverse momentum in the $K^*$ ($\phi$) meson,
respectively. It turns out that this effect increases $R_L$ by
about 5\%, and is negligible. It also decreases the relative
strong phases $\phi_\parallel$ and $\phi_\perp$ a bit.

\subsection{Rescattering Effect}

It has been proposed to explain the $B\to\phi K^*$ polarization
data through the rescattering effect \cite{CDP,LLNS,CCS},
\begin{eqnarray}
B \to D_s^{(*)} D^{(*)} \to \phi K^*\;.
\end{eqnarray}
The motivation is that the longitudinal polarization fraction of
the intermediate states $D_s^{*} D^{*}$, as low as 0.5, might
propagate into the final state $\phi K^*$. First, the massive $B$
meson can decay into $\phi K^*$ through many intermediate states.
The analysis in \cite{CDP,LLNS,CCS} was restricted to only a few
channels, and likely to be model-dependent \cite{Ligeti04}. The
truncation of the higher intermediate states in this kind of
analyses has been criticized \cite{W}. Second, if this mechanism
works for the $B\to\phi K^*$ modes, it will also work for
$B\to\rho K^*$, which involve the same intermediate states. As
obtained in \cite{CCS}, $R_L$ of both the $B^+\to\rho^+ K^{*0}$
and $B^+\to\rho^0 K^{*+}$ decays are as low as 0.6. This
observation is expected, since the additional tree amplitudes in
the latter can not change $R_L$ very much. However, the data in
Table~\ref{tab:tab1} indicate $R_L\sim 0.96$ for the $B^+\to\rho^0
K^{*+}$ decay. In other words, the $B^+\to\rho^0 K^{*+}$
polarization data, obeying the naive counting rules, have strongly
constrained the rescattering effect.
Third, the $D_s^*D$ and $D_s D^*$ intermediate states,
contributing to the $P$-wave component, could affect the
perpendicular polarization of the $B\to\phi K^*$ decays.
Unfortunately, there exists a strong cancellation among these two
channels due to the CP and SU(3) (CPS) symmetries \cite{CCS}. The
$D_s^* D^*$ intermediate state survives the CPS symmetry, which,
however, exhibits a vanishing $R_\perp$ as in Eq.~(\ref{arr}).
Therefore, the rescattering effect leads to the pattern,
\begin{eqnarray}
R_L\sim R_\parallel \gg R_\perp\;, \label{res}
\end{eqnarray}
contrary to the observed approximate equality $R_\parallel\approx
R_\perp$.

Furthermore, it has been known that the $B\to KK$ decays are
sensitive to rescattering effects. The $B\to KK$ branching ratios
measured recently well agree with the PQCD predictions \cite{CL00}
as shown in Table~\ref{tab3}, leaving very limited room for the
rescattering effect. Note that no theoretical errors were
presented in \cite{CL00}, since the detailed investigation of
uncertainties in the PQCD approach was available only after
Ref.~\cite{TLS}. Roughly speaking, the theoretical errors on PQCD
predictions for branching ratios of two-body charmless $B$ meson
decays are about 30\%. Viewing the contradiction of
Eq.~(\ref{res}) to the $B\to\phi K^*$ polarization data, and the
constraints from the measured $B^+\to\rho^0 K^{*+}$ polarizations
and from the measured $B\to KK$ branching ratios, we intend to
conclude that the rescattering effect is not a satisfactory
resolution to the polarization puzzle.

\begin{table}[ht]
\begin{center}
\begin{tabular}{c c c}\hline \hline
Branching Ratio&PQCD& Babar \cite{Bar080}\\
\hline $B(B^+\to K^+K^0)$&$1.65\times
10^{-6}$&$(1.45^{+0.53}_{-0.46}\pm
0.11)\times 10^{-6}$\\
$B(B^0\to K^0{\bar K}^0)$&$1.75\times 10^{-6}$&$(1.19^{+0.40}_{-0.35}\pm 0.13)\times 10^{-6}$\\
\hline \hline
\end{tabular}
\end{center}
\caption{PQCD predictions for the CP-averaged $B\to KK$ branching
ratios and the data. }\label{tab3}
\end{table}

\subsection{Magnetic Penguin}

\begin{figure}[t]
\centerline{
\includegraphics[width=13cm]{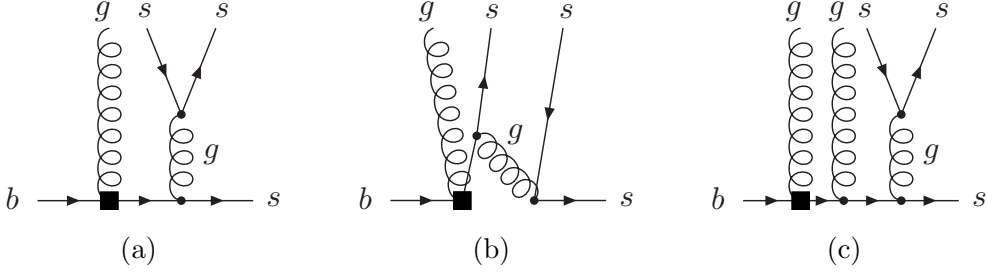}
} \caption{Some diagrams from the magnetic penguin which
contribute to the transverse polarization fraction of the
$B\to\phi K^*$ decays.} \label{fig4}
\end{figure}

If the $B\to\rho K^*$ data can be understood in the Standard Model
by means of the penguin annihilation and nonfactorizable
contributions, and only the $B\to\phi K^*$ decays exhibit an
anomaly, it is natural to look for a unique mechanism for the
latter. Such a mechanism, the $b\to sg$ transition, has been
proposed in \cite{HN}. The novel idea is that the transversely
polarized gluon from the transition propagates into the $\phi$
meson, enhancing the transverse polarization amplitudes. The
relevant matrix element was then parameterized in terms of a
dimensionless free parameter $\kappa$. Assuming this parameter to
be $\kappa\sim -0.25$, the authors of \cite{HN} claimed that the
$B\to\phi K^*$ polarization data could be accommodated within the
Standard Model. Similarly, varying a free parameter to fit the
data can not be conclusive, and a reliable estimate of the
$\kappa$ value is necessary. As pointed out in \cite{HN}, the same
mechanism also contributes to the $B\to\omega K^*$ decays, and
small $R_L\sim 0.5$ have been predicted. Therefore, the
measurement of the $B\to\omega K^*$ polarizations will impose a
stringent test on this proposal in the future. Note that PQCD
postulates, contrary to \cite{HN}, that $R_L$ of the $B\to\omega
K^*$ decays are as large as those of $B\to\rho^0 K^*$.

Besides the above experimental discrimination, we shall estimate
the order of magnitude of $\kappa$ in the framework of FA,
following the method in \cite{CL0307}. The weak effective
Hamiltonian contains the $b\to sg$ transition,
\begin{eqnarray}
-\frac{G_F}{\sqrt{2}}V_{ts}^*V_{tb}C_{8g}O_{8g}\;,
\end{eqnarray}
with the magnetic penguin operator,
\begin{eqnarray}
O_{8g}=\frac{g}{8\pi^2}m_b{\bar
s}_i\sigma_{\mu\nu}(1+\gamma_5)T_{ij}^aG^{a\mu\nu}b_j\;,
\end{eqnarray}
$i$, $j$ being the color indices. The picture described in
\cite{HN} is displayed in Fig.~\ref{fig4}: one or more collinear
gluons, emitted from the $B\to K^*$ form factor, produce the $s$
and $\bar s$ quarks in the color-octet state. They, together with
the transversely polarized gluon from the $b\to sg$ transition,
fragment into the color-singlet transversely polarized $\phi$
meson. According to this picture, we introduce three-parton
distribution amplitudes to absorb the nonperturbative dynamics
associated with the $\phi$ meson. Another diagram, in which the
transversely polarized gluon produces the $s$ and $\bar s$ quarks
and the collinear gluon flows into the $\phi$ meson directly, does
not contribute, since the transverse polarization of the $\phi$
meson is mainly carried by its gluonic parton.

We first argue that the leading picture in Fig.~\ref{fig4}(a),
where the collinear gluon attaches the $s$ quark from the $b\to
sg$ transition, diminishes due to the $G$-parity. The
corresponding three-parton distribution amplitudes are defined via
the matrix elements \cite{BBKT},
\begin{eqnarray}
& &\langle \phi(P_3,\epsilon^*_3(T))|\bar
s(-z)gG_{\mu\nu}(vz)\gamma_\alpha
s(z)|0\rangle\nonumber\\
& &=-iP_{3\alpha}\left[P_{3\mu}\epsilon^*_{3\nu}(T)-
\epsilon^*_{3\mu}(T)P_{3\nu}\right] f_{3\phi}^V {\tilde
V}(v,P_3\cdot z)\;,\label{mat1}\\
& &\langle\phi(P_3,\epsilon^*_3(T))|\bar s(-z)g\tilde
G_{\mu\nu}(vz)\gamma_\alpha\gamma_5
s(z)|0\rangle\nonumber\\
& &=P_{3\alpha}\left[P_{3\nu}\epsilon^*_{3\mu}(T)
-P_{3\mu}\epsilon^*_{3\nu}(T)\right] f_{3\phi}^A {\tilde
A}(v,P_3\cdot z)\;,\label{mat2}
\end{eqnarray}
with the dual gluon field strength tensor $\tilde
G_{\mu\nu}=\epsilon_{\mu\nu\rho\sigma}G^{\rho\sigma}/2$. Other
three-parton distribution amplitudes, irrelevant to the discussion
below, are not quoted here. The Fourier transformation of the
distribution amplitude $\tilde V$ gives
\begin{eqnarray}
{\tilde V}(v,P_3\cdot z)=\int [dx]\exp[iP_3\cdot z(x_{\bar
s}-x_s+vx_g)]V(x_s,x_{\bar s},x_g)\;,\label{four}
\end{eqnarray}
with $x_s$, $x_{\bar s}$, and $x_g$ being the momentum fractions
carried by the $s$ quark, the ${\bar s}$ quark, and the gluon,
respectively, and the integration measure,
\begin{eqnarray}
\int [dx]\equiv \int_0^1 dx_{\bar s}\int_0^1  dx_s\int_0^1
dx_g\delta\left(1-\sum_i x_i\right)\;.
\end{eqnarray}
The Fourier transformation of $\tilde A$ is defined in the same
way. The three-parton twist-3 vector meson distribution amplitudes
have been also studied in \cite{IH}. The asymptotic models of $V$
and $A$ have been parameterized as \cite{BBKT,CZ84,SVZ}
\begin{eqnarray}
V(x_s,x_{\bar s},x_g)&=&5040(x_s-x_{\bar s})x_sx_{\bar
s}x_g^2\;,\label{vda}\\
A(x_s,x_{\bar s},x_g)&=&360x_sx_{\bar
s}x_g^2\left[1+\omega_{1,0}^A\frac{1}{2}(7x_g-3)\right] \;,
\end{eqnarray}
with the shape parameter $\omega_{1,0}^A=-2.1$. The antisymmetry
(symmetry) of $V$ ($A$) between the exchange of $x_s$ and $x_{\bar
s}$ is a consequence of the $G$-parity transformation in the SU(3)
limit \cite{BBKT}. The constant $f_{3\phi}^V$ is chosen such that
$V$ is normalized according to
\begin{eqnarray}
\int [dx](x_s-x_{\bar s})V(x_s,x_{\bar s},x_g)=1\;.\label{vdan}
\end{eqnarray}

We factorize the matrix element in Eq.~(\ref{mat1}) and the $B\to
K^*$ transition form factor out of Fig.~\ref{fig4}(a). The
remaining part is the hard kernel, which must be symmetric under
the exchange of $x_s$ and $x_{\bar s}$. Therefore,
Fig.~\ref{fig4}(a), written as the convolution of the symmetric
hard kernel with the antisymmetric three-parton distribution
amplitude $V$, vanishes. The diagram with the collinear gluon
attaching the $b$ quark does not contribute, because its hard
kernel is also symmetric. No matter how many infrared gluons are
involved in the $s$-$\bar s$ quark pair production, there is no
contribution for the same reason. If factorizing the matrix
element in Eq.~(\ref{mat2}) out of Fig.~\ref{fig4}(a), it is easy
to find that the corresponding hard kernel vanishes, because the
$s$-$\bar s$ quark pair does not form an axial-vector current.
Therefore, we conclude that the leading diagram does not
contribute to the transverse polarization amplitudes of the
$B\to\phi K^*$ decays.

To survive the above suppressions, one has to consider subleading
diagrams such as Fig.~\ref{fig4}(b), in which both the $s$ quark
and the gluon from $b\to sg$ flow into the $\phi$ meson, or such
as Fig.~\ref{fig4}(c), in which one more infrared gluon fragments
into the $\phi$ meson. For Fig.~\ref{fig4}(b), an extra hard gluon
is necessary for producing the $s$-$\bar s$ quark pair, such that
the price to pay is the $\alpha_s$ suppression. For
Fig.~\ref{fig4}(c), four-parton distribution amplitudes are
involved, whose contribution is power-suppressed. We shall show
that the order of magnitude of $\kappa$ from Fig.~\ref{fig4}(b)
is, unfortunately, as small as 0.01, far away from $\kappa\sim
-0.25$ required by the data. For simplicity, we analyze the
contribution from the three-parton distribution amplitude $V$, and
our conclusion applies to that from $A$. Due to the lack of the
information of the four-parton twist-4 distribution amplitudes, we
can not estimate the contribution from Fig.~\ref{fig4}(c) in a
reliable way. However, it is of higher power in $m_\phi/m_B$, and
unlikely to be huge.

Insert the Fierz identity,
\begin{eqnarray}
I_{ij}I_{lk}=\frac{1}{4}(\gamma_\alpha)_{ik}(\gamma^\alpha)_{lj}
+\cdots\;,
\end{eqnarray}
where the irrelevant terms have been suppressed, and the identity
for color matrices,
\begin{eqnarray}
I_{ij}I_{lk}=2(T^b)_{ik}(T^b)_{lj} +\frac{1}{N_c}I_{ik}I_{lj}\;,
\end{eqnarray}
to change the fermion and color flows of the outgoing $s$ and
$\bar s$ quarks, respectively. Figure \ref{fig4}(b), where the
additional hard gluon attaches the $s$ quark going into the $K^*$
meson, is then factorized into
\begin{eqnarray}
{\cal M}&=&-\frac{G_F}{\sqrt{2}}V_{ts}^*V_{tb}
\frac{g^2}{8\pi^2}m_b C_{8g}\int[dx]{\rm IFT}\langle
\phi(P_3,\epsilon^*_3(T))|\bar
s(-z)gG^{a\mu\nu}(vz)T^b\gamma_\alpha
s(z)|0\rangle\nonumber\\
& & \times\frac{1}{4}tr\left[\cdots\gamma^\lambda\gamma^\alpha
\gamma_\lambda\frac{\not P_1-x_g\not P_3}{(P_1-x_gP_3)^2}
\sigma_{\mu\nu}(1+\gamma_5)\cdots\right] \frac{
2tr(T^cT^bT^cT^a)}{(P_2+x_{\bar s}P_3)^2}\;,
\end{eqnarray}
where IFT means the inverse Fourier transformation. The indices
$\lambda$ denote the hard gluon vertices, and the dots in the
trace represent the Feynman rules associated with the $B\to K^*$
form factor.
Employing $2tr(T^cT^bT^cT^a)=-\delta^{ab}/(2N_c)$,
Eq.~(\ref{mat1}) and Eq.~(\ref{four}), the above expression
becomes
\begin{eqnarray}
{\cal M}&=&-\frac{G_F}{\sqrt{2}}V_{ts}^*V_{tb}
\frac{g^2}{8\pi^2N_c}m_b C_{8g}f_{3\phi}^V\int[dx]V(x_s,x_{\bar
s},x_g)
\nonumber\\
& & \times\frac{1}{4}tr\left[\cdots\gamma^\lambda\not P_3
\gamma_\lambda\frac{\not P_1-x_g\not P_3}{(P_1-x_gP_3)^2}
i\sigma_{\mu\nu}(1+\gamma_5)\cdots\right]
\frac{P_3^{\mu}\epsilon^{*\nu}_3(T)}{(P_2+x_{\bar s}P_3)^2}\;.
\end{eqnarray}
Neglecting the light meson masses, assuming $m_b\approx m_B$, and
working out the product of the Dirac matrices in the trace, we
derive
\begin{eqnarray}
{\cal M}&=&-\frac{G_F}{\sqrt{2}}V_{ts}^*V_{tb}
\frac{\alpha_s}{4\pi N_c}C_{8g}\frac{f_{3\phi}^V}{m_B}
\int[dx]\frac{V(x_s,x_{\bar s},x_g)}{x_{\bar
s}(x_s+x_{\bar s})}\nonumber\\
& &\times \langle K^{*-}(P_2,\epsilon^{*}_2(T))|\bar
si\sigma_{\mu\nu}(1+\gamma_5)b|B^-(P_1)\rangle
P_3^\nu\epsilon_3^{*\mu}(T) \;.
\end{eqnarray}

Comparing the above expression with Eq.~(4) in \cite{HN}, the
parameter $\kappa$ is given by
\begin{eqnarray}
\kappa=\frac{\alpha_s}{4\pi N_c}\zeta_{3\phi}^V
\int[dx]\frac{V(x_s,x_{\bar s},x_g)}{x_{\bar s}(x_s+x_{\bar
s})}\;,
\end{eqnarray}
with $\zeta_{3\phi}^V\equiv f_{3\phi}^V/(f_\phi m_\phi)$. For the
values $\alpha_s=0.4$ and $\zeta_{3\phi}^V=0.013$ \cite{BBKT}, and
the model distribution amplitude in Eq.~(\ref{vda}), we obtain
$\kappa\approx 0.004$. Other diagrams with the hard gluon
attaching the $b$ quark and the transversely polarized gluon can
be analyzed in a similar way, and the results are of the same
order of magnitude.
Adding these contributions leads to
\begin{eqnarray}
\kappa\approx 0.01\;.
\end{eqnarray}
Hence, we intend to conclude that the magnetic penguin is not
sufficient to resolve the $B\to\phi K^*$ puzzle.

\section{CONCLUSION}

In this paper we have investigated most of the $B\to VV$ modes
carefully. Our observation is that the $B\to VV$ modes can be
classified into four categories. For the tree-dominated decays,
the polarization fractions are basically determined by kinematics.
The upper part of Table~\ref{tab:tab2} can be understood by
kinematics in the heavy-quark limit. The longitudinal polarization
fractions follow the mass hierarchy among the $D_s^{*}$, $D^{*}$
and $\rho$ ($K^*$) mesons. The lower part of Table~\ref{tab:tab2}
can be understood by kinematics in the large-energy limit. We
always have $R_L\sim 1$ for the decays into two light vector
mesons. It has been found that the above simple kinematic
estimates in the heavy-quark and large-energy limits are robust
under subleading corrections. For this part, we have analyzed the
next-to-leading-power corrections to the universal IW function,
the nonfactorizable contributions, the two-parton twist-4
contributions, and part of next-to-leading-order contributions
from the charm-quark loop, all of which are negligible. That is,
QCD dynamics plays only a minor role for the polarizations of the
tree-dominated decays.

As a byproduct, we have predicted the longitudinal polarization
fractions of the $B^+\to (D_s^{*+}, D^{*+})\rho^0$ modes in the
large-energy limit using FA, and found $R_L\sim 0.7$. We have also
calculated the polarization fractions of the $B^0\to {\bar
D}^{*0}D^{*0}$ decay explicitly in the PQCD approach based on
$k_T$ factorization theorem. The result $R_L\sim 1$ is quite
different from $R_L$ of other $B\to D^*D^*$ modes, since it is
dominated by the nonfactorizable $W$-exchange topology. The above
predictions can be confronted with the future data.

For the penguin-dominated modes, the polarization fractions can
deviate from the naive counting rules based on kinematics, because
of the important annihilation contribution from the $(S-P)(S+P)$
operators. This mechanism explains the third category listed in
the lower part of Table~\ref{tab:tab1}: $R_L$ can decrease to 0.75
for the $B^+\to\rho^+ K^{*0}$ mode. Adding the tree contribution,
$R_L$ of the $B^+\to\rho^0 K^{*+}$ decay can go up to about 0.9.
We have postulated from the viewpoint of PQCD that the $B\to\omega
K^*$ decays also belong to the third category, and should show
$R_L$ similar to those of $B\to\rho^0 K^*$. All the above three
categories can be accommodated within the Standard Model. Only the
fourth category, the $B\to\phi K^*$ decays, can not. They are
dominated by the penguin contribution, but their $R_L\sim 0.5$ are
much lower than 0.75. We have carefully analyzed the various
mechanism proposed in the literature to resolve this anomaly, and
concluded that none of them is satisfactory. Therefore, the
$B\to\phi K^*$ polarization data remain as a puzzle. However, we
emphasize that we are not claiming a signal of new physics, since
the complicated QCD dynamics in the $B\to VV$ decays has not yet
been fully explored. For example, a smaller $B\to K^*$ form factor
$A_0$ could decrease $R_L$ significantly \cite{L0411}.

\vskip 1.0cm We thank C. Bauer, I.I. Bigi, P. Chang, C.H. Chen,
K.F. Chen, H.Y. Cheng, C.K. Chua, W.S. Hou, Y.Y. Keum, Z. Ligeti,
M. Nagashima, D. Pirjol, A.I. Sanda, and I. Stewart for useful
discussions. This work was supported by the National Science
Council of R.O.C. under Grant No. NSC-93-2112-M-001-014, by the
Taipei Branch of the National Center for Theoretical Sciences of
R.O.C., and by the Grants-in-aid from the Ministry of Education,
Culture, Sports, Science and Technology, Japan under Grant No.
14046201. HNL acknowledges the hospitality of Department of
Physics, Tohoku University, where this work was initiated.

\end{document}